\documentclass[11pt,a4paper]{article} 
\usepackage{jheppub}

\usepackage{epsfig}
\usepackage{latexsym}
\usepackage{graphics}
\usepackage{graphicx}
\usepackage{epstopdf}  
\usepackage[justification=RaggedRight]{caption}
\usepackage{amssymb}
\usepackage{subfig}
\usepackage{color}
\usepackage{multirow}
\usepackage{color}
\usepackage{rotating}

\usepackage{hyperref} 

\usepackage{hyperref}
\hypersetup{
    colorlinks,%
    citecolor=blue,%
    filecolor=blue,%
    linkcolor=blue,%
    urlcolor=blue
}

\def\be{\begin{equation}}
\def\ee{\end{equation}}

\newcommand{\vev}[1]{\left\langle #1\right\rangle}

\newcommand{\bwt}{\begin{widetext}}
\newcommand{\ewt}{\end{widetext}}
\newcommand{\bdm}{\begin{displaymath}}
\newcommand{\edm}{\end{displaymath}}
\newcommand{\bea}{\begin{eqnarray}}
\newcommand{\eea}{\end{eqnarray}}

\def\eq#1{{Eq.~(\ref{#1})}}
\def\eqs#1#2{{Eqs.~(\ref{#1})--(\ref{#2})}}
\def\fig#1{{Fig.~\ref{#1}}}

\def\Table#1{{Table~\ref{#1}}}

\def\sect#1{{Sect.~\ref{#1}}}

\def\app#1{{Appendix~\ref{#1}}}

\def\vev#1{\left\langle #1\right\rangle}
\def\abs#1{\left| #1\right|}

\begin{document}

\title{Minimal Flavour Violation and 
Neutrino Masses without R-parity}

\author[a]{Giorgio Arcadi,} 
\author[b] {Luca Di Luzio,}
\author[c]{ Marco Nardecchia}

\affiliation[a]{SISSA/ISAS and INFN, Sezione di Trieste, \\
Via Bonomea 265, I-34136 Trieste, Italy} 
\affiliation[b]{Institut f\"{u}r Theoretische Teilchenphysik,
Karlsruhe Institute of Technology (KIT), \\ D-76128 Karlsruhe, Germany} 
\affiliation[c]{CP3-Origins and Danish Institute for Advanced Study (DIAS), University of Southern Denmark,\\ 
Campusvej 55, DK-5230, Odense M, Denmark}

\emailAdd{arcadi@sissa.it}
\emailAdd{diluzio@particle.uni-karlsruhe.de}
\emailAdd{nardecchia@cp3-origins.net}

\abstract{
We study the extension of the Minimal Flavour Violation (MFV) hypothesis to the MSSM without R-parity. 
The novelty of our approach lies in the observation that supersymmetry enhances the global symmetry of the kinetic term 
and in the fact that we consider as irreducible sources of the flavour symmetry breaking all the couplings of the superpotential including the R-parity violating ones. 
If R-parity violation is responsible for neutrino masses, our setup can be seen as an extension of MFV to the lepton sector. 
We analyze two patterns based on the non-abelian flavour symmetries $SU(3)^4 \otimes SU(4)$ and $SU(3)^5$. 
In the former case the total lepton number and the lepton flavour number are broken together, 
while in the latter the lepton number can be broken independently by an abelian spurion, 
so that visible effects and peculiar correlations can be envisaged in flavour changing charged lepton decays like $\ell_i \rightarrow \ell_j \gamma$.
}

\keywords{}

\maketitle

\section{Introduction}

The flavour problem can be viewed as the clash between the theoretical expectation of New Physics (NP) at the 
TeV scale and the experimental observations in Flavour Changing Neutral Current (FCNC) processes 
which severely constrain the scale $\Lambda_{\rm{NP}}$ of the NP beyond the $10^4$ TeV 
domain (for a review see e.g.~Ref.~\cite{Isidori:2010kg}). 
If we insist in keeping $\Lambda_{\textrm{NP}} \approx$ TeV for naturalness, then we have to conclude that 
the flavour structure of the NP is highly non-generic. 

The Minimal Flavour Violation (MFV) hypothesis~\cite{Chivukula:1987py,Hall:1990ac,Buras:2000dm,D'Ambrosio:2002ex} 
is a powerful organizing principle which states that 
the sources of flavour symmetry breaking of the NP are aligned to the Standard Model (SM) Yukawas.  
This ansatz provides an automatic suppression of the NP contribution to the flavour violating observables and 
thus a solution of the aforementioned flavour problem (see for instance Ref.~\cite{Bona:2007vi,Hurth:2008jc}).

In particular the MFV hypothesis can be formulated as a symmetry principle: 
in absence of the Yukawa couplings the global symmetry of the SM 
is that of the gauge invariant kinetic terms. 
This flavour symmetry or a subgroup can be formally be restored by promoting the Yukawa couplings 
to spurions with definite transformation properties under the flavour group and the same holds in any extension of the SM.

While in the quark sector the MFV ansatz unambiguously relates 
the sources of flavor breaking beyond the SM to the quark masses and the CKM matrix, 
the situation in the lepton sector is less straightforward. This is namely due to the introduction 
of new sources of flavour breaking related to neutrino masses and, since the mechanism itself 
generating neutrino masses is unknown, several scenarios can be envisaged. 
Starting from Ref.~\cite{Cirigliano:2005ck} many formulations of Minimal Lepton Flavour Violation (MLFV) 
have been proposed and analyzed~\cite{Cirigliano:2006su,Cirigliano:2006nu,Davidson:2006bd,Branco:2006hz,Gavela:2009cd,Filipuzzi:2009xr,Alonso:2011jd}.

In this work we focus on a particular realization of MFV in the in the context of the Minimal Supersymmetric SM (MSSM) without R-parity 
(for a review see e.g.~Ref.~\cite{Barbier:2004ez}).
Our analysis is moved by two simple observations about the MSSM:
\begin{enumerate}
\item The largest group of unitary transformations commuting with the gauge group  
is enhanced with respect to the SM case. This is namely 
$U(3)_{\hat{q}} \otimes U(3)_{\hat{u}^c} \otimes U(3)_{\hat{d}^c} \otimes U(3)_{\hat{e}^c} \otimes U(4)_{\hat{L}} \otimes U(1)_{\hat{h}_u}$, 
where the presence of the $U(4)_{\hat{L}}$ factor is due to the fact that the superfields $\hat{\ell}$ and $\hat{h}_d$ have the same quantum numbers \cite{Hall:1983id,Hempfling:1995wj,Nilles:1996ij,Nardi:1996iy,Chun:1999bq,Davidson:2000ne}. 
 
\item The MSSM (without right handed neutrinos) has already all the degrees of freedom sufficient to generate 
neutrino masses and mixings through R-parity Violating (RPV) 
interactions~\cite{Hall:1983id,Lee:1984kr,Lee:1984tn,Dawson:1985vr,Banks:1995by,Nardi:1996iy,Hirsch:2000ef,Hirsch:2004he,Bajc:2010qj}.
\end{enumerate}
Thus the aim of our work is to generalize the MFV expansion of the soft terms 
in the presence of the enlarged flavor symmetry and to include the RPV couplings as the original sources of flavor breaking. 
Such a MFV expansion is in principle expressed in terms of many free parameters. However, 
motivated by our second observation, we connect the RPV spurions with the neutrino sector observables, 
requiring that the flavour symmetry is broken in the minimal way compatible with the pattern of neutrino 
masses. 
In particular we will study the impact of such a framework on LFV processes, thus 
providing an alternative scenario of MLFV.  

Notice that the MSSM without R-parity contains 
a large set of lepton number violating parameters, 
making 
the connection between neutrino masses and the soft terms still ambiguous.   
Thus extra assumptions on the orientation 
of the RPV spurions in the flavour space are needed.
 
Our approach towards the R-parity differs from that of Refs.~\cite{Nikolidakis:2007fc,Csaki:2011ge} 
in the fact that we do not aim at an explanation of the smallness of the RPV couplings.
We simply treat them, in a more democratic way, on the same ground of all the other couplings of the superpotential.
Remarkably, the values of the RPV couplings needed in order to fit neutrino masses 
are of the same order of magnitude of the SM Yukawas of the first and second families~\cite{Allanach:2003eb,Ellis:1998rj}. 

In the following we analyze two symmetry patterns based on the groups
$SU(3)^4 \otimes SU(4)$ and $SU(3)^5 \otimes U(1)_L \otimes U(1)_B$. 
With the former flavour symmetry the breaking scale of lepton number 
is related to that of lepton flavour violation (LFV), 
thus implying small effects in LFV physics. On the other hand the 
latter flavour symmetry allows to break the lepton number independently by means of an abelian spurion, 
so that visible effects are in principle achievable. 
We finally study the correlations 
among the flavour changing charged lepton decays $\ell_i \rightarrow \ell_j \gamma$.

\section{Minimal Flavour Violation without R-parity}
\label{flavgroupMSSM}

The starting point of the MFV idea is based on the observation that
the largest group of unitary transformations commuting with the SM gauge group is
\be
\label{kineticSM}
G_{\rm{kin}}^{\rm{SM}} = U(3)_{q} \otimes U(3)_{u^c} \otimes U(3)_{d^c} \otimes U(3)_{e^c} \otimes U(3)_{\ell} \otimes U(1)_{h} \, .
\ee
This corresponds to the global symmetry of the gauge invariant kinetic term of the SM fields 
\be
\Phi = \left( q_i, u^c_i, d^c_i, e^c_i, \ell_i, h \right) \, ,
\ee
with $i$ spanning over the three families. Notice that $\ell_i$ and $h$ have the same gauge quantum numbers and only the Lorentz structure 
prevents the global symmetry of the kinetic term from being larger. 

On the other hand the situation in the MSSM is qualitatively different since the 
supersymmetrization of the SM spectrum restore the symmetry between scalars and fermions, thus enhancing the global symmetry of the kinetic term.  

In order to make apparent this enhancement it is useful to define a generalized lepton multiplet $\hat{L}_{\alpha} = (\hat{\ell}_i, \hat{h}_d)$ 
and rewrite the set of chiral superfields of the MSSM in the following way
\be
\hat{\Phi}= \left( \hat{q}_i, \hat{u}^c_i, \hat{d}^c_i,\hat{e}^c_i, \hat{L}_{\alpha}, \hat{h}_u \right) \, , 
\ee
where a second Higgs doublet is introduced in order to ensure anomaly cancellation. 
Then the global symmetry of the kinetic term
\begin{equation}
\int d^4  \theta \ \hat{\Phi}^{\dagger} e^{2 g \hat{V}} \hat{\Phi} 
\end{equation}
turns out to be 
\be
\label{kineticMSSM}
G_{\rm{kin}}^{\rm{MSSM}} = U(3)_{\hat{q}} \otimes U(3)_{\hat{u}^c} \otimes U(3)_{\hat{d}^c} \otimes U(3)_{\hat{e}^c} \otimes U(4)_{\hat{L}} \otimes U(1)_{\hat{h}_u} \, .
\ee
Notice that this holds irrespectively of the fact that R-parity is or not an exact symmetry of the full MSSM lagrangian. 

We can decompose $G_{\rm{kin}}^{\rm{MSSM}}$ in abelian and non-abelian factors 
and identify a linear combination of the six $U(1)$ generators with the SM hypercharge. 
Then we define the generalized flavour group of the MSSM as
\be
G_{F}= SU(3)_{\hat{q}} \otimes SU(3)_{\hat{u}^c} \otimes SU(3)_{\hat{d}^c} \otimes SU(3)_{\hat{e}^c} \otimes SU(4)_{\hat{L}} \, ,                  
\ee
while the abelian factors can be rearranged in the following way
\be
G_{A} = U(1)_{\hat{u}^c} \otimes U(1)_{\hat{d}^c} \otimes U(1)_{\hat{e}^c} \otimes U(1)_{\hat{L}} \otimes U(1)_{B} \, ,                  
\ee
where $B$ is the baryon number. 
$G_F$ and $G_A$ are explicitly broken by the most general MSSM superpotential and soft lagrangian. 

Since the MSSM has many sources of flavour violation it is useful to have a rationale in order to select the origin of this 
breaking. 
Let us imagine that the flavour symmetry is broken at the scale $\Lambda_F$ 
by some unknown mechanism. 
Then, if the breaking of SUSY is due to a flavour universal mechanism (like in gauge mediation \cite{Giudice:1998bp})  
and the scale of mediation $M$ is smaller than $\Lambda_F$,  the flavor structure of the soft terms will be related to the breaking of the flavor symmetry in the supersymmetric sector.

Having in mind such a MFV framework we \emph{assume} that the original source of flavour violation is given by the 
the couplings of the most general MSSM superpotential 
\begin{equation}
\label{susyW}
W= Y_U^{ij} \hat{q}_i \hat{u}^c_j \hat{h}_u + Y_D^{\alpha ij} \hat{L}_{\alpha} \hat{q}_i \hat{d}^c_j 
+ \tfrac{1}{2} Y_E^{\alpha \beta i} \hat{L}_{\alpha} \hat{L}_{\beta} \hat{e}^c_i + \mu^{\alpha} \hat{h}_u \hat{L}_{\alpha}  
+ \tfrac{1}{2}  (\lambda^{''} )^{ijk}  \hat{u}^c_i   \hat{d}^c_j  \hat{d}^c_k  \, ,
\end{equation}
where the gauge structure has been omitted for simplicity.
Notice also the antisymmetry of the couplings $Y_E^{\alpha \beta i} = - Y_E^{\beta \alpha i}$ and $(\lambda^{''} )^{ijk} = - (\lambda^{''} )^{ikj}$.

In order to formally restore the invariance with respect to the flavor group 
we treat the couplings in~\eq{susyW} as spurions, with quantum numbers 
under $G_F$:
\bea
\label{YUspu}
Y_U &\sim& (\bar{3},\bar{3},1,1,1)
\\
\label{YDspu}
Y_D &\sim& (\bar{3},1,\bar{3},1,\bar{4})
\\
\label{YEspu}
Y_E &\sim& (1,1,1,\bar{3},6)
\\
\label{muspu}
\mu &\sim& (1,1,1,1,\bar{4})
\\
\label{lambdasspu}
\lambda^{''} &\sim& (1,\bar{3},3,1,1)
\, ,
\eea
where our conventions are such that each chiral superfield in $\hat{\Phi}$ transforms according to the fundamental representation of 
the corresponding group factor in $G_F$.

Following the MFV principle we can expand the soft terms (cf.~\app{appnotation} for the notation) by means of the spurions in~\eqs{YUspu}{lambdasspu}
\begin{equation}
\label{MFVexpSU4}
\begin{array}{rcl}
\left( \tilde{m}^2_q \right)^{i}_j & = & \tilde{m}^2 \left( c_q  \delta^{i}_j + d^{1}_{q} \ Y_U^{ik} \left(Y_U^* \right)_{jk} 
+ d^{2}_{q} \ Y_D^{\alpha ik} \left(Y_D^* \right)_{\alpha jk} \right)  \\
\left( \tilde{m}^2_{u^c} \right)^{i}_j & = & \tilde{m}^2 \left( c_{u^c}  \delta^{i}_j +  d^{1}_{u^c} \ Y_U^{ki} \left(Y_U^* \right)_{kj} 
+ d^{2}_{u^c} (\lambda^{''} )^{ikl} (\lambda^{''\ast} )_{jkl} \right) \\
\left( \tilde{m}^2_{d^c} \right)^{i}_j & = & \tilde{m}^2 \left( c_{d^c}  \delta^{i}_j + d^{2}_{d^c} \ Y_D^{\alpha k i} \left(Y_D^* \right)_{\alpha k j} 
+ d^{2}_{d^c} (\lambda^{''} )^{kil} (\lambda^{''\ast} )_{kjl} \right)  \\

\left( \tilde{m}^2_{e^c} \right)^{i}_j & = & \tilde{m}^2 \left( c_{e^c}  \delta^{i}_j + d^{1}_{e^c} \ Y_E^{\alpha \beta i} \left(Y_E^* \right)_{\alpha \beta j} \right)  \\
\left( \tilde{m}^2_{L} \right)^{\alpha}_{\beta} & = & \tilde{m}^2 \left( c_{L}  \delta^{\alpha}_{\beta} + d^{1}_{L} \ Y_E^{\alpha \gamma k} \left(Y_E^* \right)_{\beta \gamma k} 
+ d^{2}_{L} \ Y_D^{\alpha k l} \left(Y_D^* \right)_{\beta k l}
+ d^3_L \, \mu^{\alpha} \mu^*_{\beta} / |\mu|^2  \right)\\

B^{\alpha}&= & \tilde{m}^2 \left( c_{B} \, \mu^{\alpha} / |\mu| + d^1_B \ Y_D^{\alpha k l} (Y_D^*)_{\beta k l} \, \mu^{\beta} / |\mu| 
+ d^2_B \ Y_E^{\alpha \beta k} (Y_E^*)_{\gamma \beta k} \, \mu^{\gamma} / |\mu|  \right) \\

A^{ij}_U &=& A \left( c_{A_U} Y^{ij}_U + d^1_{A_U} Y^{kj}_U \left( Y_D^* \right)_{\alpha k l} Y^{\alpha i l}_D + d^2_{A_U} Y^{ik}_U ( \lambda^{''*} )_{k l m } ( \lambda^{''} )^{j l m} \right. \\
& & \left.  + d^3_{A_U} Y^{ik}_U \left( Y_U^* \right)_{l k} Y^{l j}_U \right) \\

A^{\alpha ij}_D &=& A \left( c_{A_D} Y^{\alpha ij}_D 
+  d^1_{A_D} Y^{\alpha kj}_D \left(Y_U^* \right)_{kl} Y_U^{il}  
+  d^2_{A_D} Y^{\beta ij}_D \left(Y_E^* \right)_{\beta \gamma k} Y_E^{\alpha \gamma k} \right. \\
& & + d^3_{A_D} Y^{\alpha ik}_D ( \lambda^{''*} )_{l k m } ( \lambda^{''} )^{l j m} + d^4_{A_D} Y^{\alpha il}_D  (Y_D^*)_{\gamma k l} (Y_D)^{\gamma k j} \\
& & 
+ d^5_{A_D} Y^{\alpha k j}_D  (Y_D^*)_{\gamma k l} (Y_D)^{\gamma i l}  
+ d^6_{A_D} Y^{\alpha kl}_D  (Y_D^*)_{\gamma k l} (Y_D)^{\gamma i j} \\
& & \left.
+ d^7_{A_D} Y^{\beta ij}_D \mu^*_{\beta} \mu^{\alpha} / |\mu|^2  \right)\\

A^{\alpha \beta i}_E &=& A \left( c_{A_E} Y^{\alpha \beta i}_E  
+  d^1_{A_E} Y^{[\alpha \gamma i} _E    \left(Y^*_D \right)_{\gamma k l}  Y^{\beta] k l} _D 
+d^2_{A_E} Y^{\alpha \beta k} _E    \left(Y^*_E \right)_{\gamma \delta k}  Y^{\gamma \delta i} _E \right. \\
& & \left. + d^3_{A_E} Y^{[\alpha \gamma k} _E    \left(Y^*_E \right)_{\gamma \delta k}  Y^{\beta] \delta i} _E 
+ d^4_{A_E} Y^{ [\alpha \gamma i}_E \mu^{*}_{\gamma} \mu^{\beta]} / |\mu|^2 \right) \\

A_{\lambda^{''}}^{ijk} &=&  A  \left( c_{A_{\lambda^{''}}}  (\lambda^{''})^{ijk} 
+ d^1_{A_{\lambda^{''}}}  (\lambda^{''})^{ljk} (Y_U^*)_{ml} (Y_U)^{mi} \right. \\ 
& &  +   d^2_{A_{\lambda^{''}}}  (\lambda^{''})^{i[jl} (Y_D^*)_{\alpha m l} (Y_D)^{\alpha m k]} \\
& & \left.
+ d^3_{A_{\lambda^{''}}}  (\lambda^{''})^{i[jm} (\lambda^{''*})_{lnm} (\lambda^{''})^{lnk]} 
+  d^4_{A_{\lambda^{''}}}  (\lambda^{''})^{imn} (\lambda^{''*})_{lmn} (\lambda^{''})^{ljk} \right) \, , \\

\end{array}
\end{equation}
where the squared brackets 
stand for anti-symmetrization 
and we also defined $\abs{\mu}^2 \equiv \sum_{\alpha=1, \dots , 4} \abs{\mu^{\alpha}}^2$.

Notice that expansion is truncated at the third order in the spurions. Actually 
the higher order terms in the spurions can be phenomenologically neglected under the assumption 
that the dimensionless couplings of the MFV expansion are of $\mathcal{O}(1)$~\cite{D'Ambrosio:2002ex,Colangelo:2008qp}.
   
In absence of R-parity all the neutral scalar components of $\hat{L}_{\alpha}$ and $\hat{h}_u$ develop a VEV 
which triggers the electroweak symmetry breaking $SU(2)_L \otimes U(1)_Y \to U(1)_Q$ \cite{Hall:1983id,Hempfling:1995wj}.  
Given the $SU(4)_{\hat{L}}$ symmetry it is always possible, without loss of generality, to redefine the $\hat{L}_{\alpha}$ superfield in 
such a way that only the fourth component acquires a VEV. 
From now on we work in such a basis and we define operatively the superfield $\hat{h}_d$ so that it 
corresponds to the component which develops 
a VEV, $\hat{h}_d \equiv \hat{L}_{4}$, while the leptons do not, $\hat{\ell}_i \equiv \hat{L}_{i}$.

Despite our notation makes explicit the underlying non-abelian flavour symmetry $SU(3)^4 \otimes SU(4)$, 
it is also useful to translate it into the more common $SU(3)^5$ language. 
This connection is provided in~\app{appnotation}.  
Then we can formally split the superpotential in~\eq{susyW} in an RPC and an RPV term
\begin{align}
& W_{RPC}= y_U^{ij} \hat{q}_i \hat{u}^c_j \hat{h}_u + y_D^{ij} \hat{h}_d \hat{q}_i \hat{d}^c_j +  y_E^{ij} \hat{h}_d \hat{\ell}_i \hat{e}^c_j 
+ \mu \, \hat{h}_u \hat{h}_d  \, , \\
& W_{RPV}= \mu^{i} \hat{h}_u \hat{\ell}_i + \tfrac{1}{2} \lambda^{ijk} \hat{\ell}_i \hat{\ell}_j \hat{e}^c_k 
+ (\lambda^{'} )^{ijk}  \hat{\ell}_i \hat{q}_j  \hat{d}^c_k +  \tfrac{1}{2}  (\lambda^{''} )^{ijk}  \hat{u}^c_i   \hat{d}^c_j  \hat{d}^c_k \, ,
\end{align}
Analogously the MFV expansion of the soft terms in~\eq{MFVexpSU4} can be easily translated in the $SU(3)^5$ language by means of the dictionary 
given in~\eq{dictionary} of~\app{appnotation}
\begin{equation}
\label{MFVexpSU3lang}
\begin{array}{rcl}
\left( \tilde{m}^2_q \right)^{i}_j & = & \tilde{m}^2 
\left( c_q \delta^i_j + d^{1}_{q} (y_U y_U^{\dagger})^i_j + d^{2}_{q} \left[ (y_D y_D^{\dagger})^i_j + (\lambda')^{lik}   \lambda'^*_{ljk} \right] \right) \\
\left( \tilde{m}^2_{u^c} \right)^{i}_j & = & \tilde{m}^2 
\left( c_{u^c} \delta^i_j +  d^{1}_{u^c}  (y_U^{\dagger}  y_U )^i_j 
+ d^{2}_{u^c} (\lambda^{''} )^{ikl} (\lambda^{''\ast} )_{jkl} \right)  \\
\left( \tilde{m}^2_{d^c} \right)^{i}_j & = & \tilde{m}^2 
\left( c_{d^c} \delta^i_j +  d^{1}_{d^c} \left[ (y_D^{\dagger} y_D)^i_j + (\lambda')^{lki}   \lambda'^*_{lkj} \right] 
+ d^{2}_{d^c} (\lambda^{''} )^{kil} (\lambda^{''\ast} )_{kjl} \right)   \\
\left( \tilde{m}^2_{e^c} \right)^{i}_j & = & \tilde{m}^2 
\left( c_{e^c} \delta^i_j +  d^{1}_{e^c} \left[ 2 (y_E^{\dagger}  y_E)^i_j +  \lambda^{lki} \lambda^*_{lkj} \right] \right)  \\
\left( \tilde{m}^2_{\ell} \right)^{i}_{j} & = & \tilde{m}^2 
\left( c_{L}  \delta^{i}_{j} + d^{1}_{L}  \left[ ( y_{E} y_E^{\dagger})^i_j + \lambda^{ilk} \lambda^*_{jlk} \right] + d^{2}_{L}  (\lambda')^{ilk} \lambda'^*_{jlk}
 + d^3_L \, \mu^{i} \mu^*_j / |\mu|^2  \right) \\
\left( \tilde{m}^2_{d} \right)^{i} & = & \tilde{m}^2 
\left( d^{1}_{L} \lambda^{ilk} (y_E^*)_{lk} + d^{2}_{L}  (\lambda')^{ilk} (y_D^{*})_{lk}
 + d^3_L \, \mu^{i} \mu^* / |\mu|^2 \right) \\
\tilde{m}^2_{h_d}& = & \tilde{m}^2 
\left(c_L +d^1_L  \  \textrm{Tr}(y_E y_E^{\dagger}) + d^2_L  \  \textrm{Tr}(y_D y_D^{\dagger}) 
+ d^3_L \, \mu \, \mu^* / |\mu|^2 \right) \\
b &= & \tilde{m}^2 \left( c_{B} \mu / |\mu| + d^1_B \left[ \textrm{Tr}(y_D y_D^{\dagger}) \, \mu / |\mu| + (y_D)^{lk} \lambda'^*_{plk} \, \mu^{p} / |\mu| \right] \right. \\
& & \left. + d^2_B \left[ \textrm{Tr}(y_E y_E^{\dagger}) \, \mu / |\mu| + (y_E)^{lk} \lambda^*_{plk} \, \mu^{p} / |\mu| \right] 
\right) \\
b^i &=&  \tilde{m}^2 \left(  c_{B} \mu^i  / |\mu|+
d^1_B \left[ (\lambda')^{ilk} (y^*_D)_{lk} \, \mu / |\mu| + (\lambda')^{ilk} \lambda'^*_{plk} \, \mu^p / |\mu| \right] \right. \\
& & \left. + d^2_B \left[  (\lambda)^{ilk} (y^*_E)_{lk} \, \mu / |\mu| + (\lambda)^{ilk} \lambda^*_{plk} \, \mu^p / |\mu| 
+ (y_E y_E^{\dagger})^i_p \, \mu^p / |\mu| \right] 
\right)\\
a^{ij}_U &=& A \left( c_{A_U} y^{ij}_U + \dots  \right) \\
a^{ij}_D &=& A \left( c_{A_D} y^{ij}_D + \dots  \right) \\
a^{ij}_E &=& A \left( c_{A_E} y^{ij}_E 
+ \ldots
\right) \\
(a_{\lambda})^{ijk} &=& A \left( c_{A_E} \lambda^{ijk} + \dots  \right) \\
(a_{\lambda^{'} })^{ijk}&=& A \left( c_{A_D} (\lambda')^{ ijk} + \dots  \right) \\
a_{\lambda^{''}}^{ijk} &=& A \left( c_{A_{\lambda^{''}}}  (\lambda^{''})^{ijk} + \dots  \right)  \, ,
\end{array}
\end{equation}
where we have omitted to expand the $a$ terms up at the third order in the spurions.

Notice that the flavor violation in the lepton sector can be linked to the amount of R-parity violation. 
For instance the RPV couplings in the expansion of $\tilde{m}^2_{\ell}$ in~\eq{MFVexpSU3lang} 
are responsible for flavour violating mass insertions leading 
to processes like $\ell_i \rightarrow \ell_j \gamma$.
Assuming that the RPV couplings are responsible for neutrino masses 
it is possible then to provide a link between LFV in the charged lepton sector 
and the neutrino observables.  

However, in absence of R-parity the MSSM superpotential is enriched by a large set of lepton number violating parameters, 
thus making the connection between neutrino masses and the soft terms somehow ambiguous.   
In order to retain some level of predictivity for the soft terms extra assumptions must be made on the orientation 
of the RPV spurions in the flavour space\footnote{Notice that this problem is present 
in many formulations of MLFV. See for instance Refs.~\cite{Cirigliano:2005ck,Davidson:2006bd,Alonso:2011jd}}.
In the next section we will exemplify this point by considering a particular model of neutrino masses in which the RPV 
spurions are oriented along the $\mu^i$ and $\lambda'^{i33}$ directions.

\section{Neutrino Masses without R-parity}

In the previous section we have formally restored the invariance under the flavour group $G_F$ by promoting 
all the supersymmetric couplings in~\eq{susyW} to spurions.
Now we want to provide the link between these spurions and the physical observables. 
Our guideline is to break the flavour group in a \emph{minimal} way, namely we consider the minimal amount of flavour breaking which is
able to reproduce the correct pattern of fermion masses and mixings.   

In the limit of massless neutrinos, the connection of the spurions with the flavour structure of the charged fermions
is straightforward. From the superpotential    
\be
W \supset Y_U^{ij} \, \hat{q}_i \hat{u}^c_j \hat{h}_u 
+ Y_D^{4ij} \, \hat{h}_d \hat{q}_i \hat{d}^c_j 
+ Y_E^{4ij} \, \hat{h}_d \hat{\ell}_{i} \hat{e}^c_j
\, ,  
\ee
we can identify the relevant spurions in terms of known physical observables, 
up to the parameter $\tan\beta \equiv v_u / v_d$.  
Indeed it is always possible to choose a basis such that 
\be
\label{RPCmassmatrices}
Y_U^{ij} = (V^\dag \hat{m}_U)^{ij} / v_u \, , 
\qquad Y_D^{4ij} = \hat{m}_D^{ij} / v_d \, , 
\qquad Y_E^{4ij} = \hat{m}_E^{ij} / v_d \, ,  
\ee
where $V$ is the CKM matrix and $\hat{m}_U$, $\hat{m}_D$, $\hat{m}_E$ are the diagonal charged-fermion 
masses. 

On the other hand the experimental evidence of neutrino masses and mixings makes clear that 
the flavour group must be further broken. 
The standard way to introduce neutrino masses in the context of supersymmetric MFV is to extend the 
field content of the MSSM by introducing three SM-singlet chiral superfields~\cite{Nikolidakis:2007fc,Csaki:2011ge} 
and thus applying the seesaw 
mechanism~\cite{Minkowski:1977sc,GellMann:1980vs,Yanagida:1979as,Glashow:1979nm,Mohapatra:1979ia,Schechter:1980gr,Schechter:1981cv}.

Remarkably the MSSM without R-parity gives the possibility of generating neutrino 
masses and mixings without the need of additional ingredients. This is the approach we pursue in this work.  
As we are going to show soon, neutrino masses are fitted by moderate small values of the R-parity violating couplings 
$\mu^i/\mu$, $\lambda$ and $\lambda'$, of $\mathcal{O}(10^{-5})$ or even larger. 
From this point of view the issue of the smallness of neutrino masses could be brought back at the same conceptual level of 
understanding the flavour structure of the charged fermions, featuring Yukawa couplings also of $\mathcal{O}(10^{-6})$ as in the case of the electron. 

The formulae for the neutrino mass matrix in terms of the RPV couplings are collected for completeness 
in~\app{RPVandNuMasses}. The leading contributions can be schematically written as
\be
\label{numassschem}
m_\nu \quad \sim \quad 
\left(\frac{M_Z}{\tilde{m}}\right)^2 \frac{\mu_i \mu_j}{\tilde{m}} \, ,
\qquad 
\frac{3 \, \lambda'^2}{8 \pi^2} \frac{\hat{m}_D^2}{\tilde{m}} \, ,
\qquad 
\frac{\lambda^2}{8 \pi^2} \frac{\hat{m}_E^2}{\tilde{m}} \, ,
\ee
where for simplicity we set $M_1 \approx M_2 \approx A \approx \mu \equiv \tilde{m}$ 
and we neglected the flavour structure of $\lambda'$ and $\lambda$. 

Finally we comment about the baryon number violating coupling $\lambda''$. 
According to our guideline at the beginning of this section, 
this coupling does not give any contribution to the construction of fermion masses and mixings and thus 
should be absent as an irreducible source of flavor breaking. 
However $\lambda''$ can still be induced by the other spurions.
If the $U(1)$ factors are part of the symmetry that we want to formally restore, 
then $\lambda''$ cannot be generated by the baryon number conserving couplings $Y_U$, $Y_E$, $Y_D$ and $\mu$. 
On the other hand if we consider only the non-abelian symmetry $SU(3)^4 \otimes SU(4)$, the coupling $\lambda''$ can be induced 
in a holomorphic way~\cite{Csaki:2011ge}:
\begin{equation}
\lambda'' \sim Y_U (Y_D)^2 (Y_E)^3 \, ,
\end{equation}
where the proper contractions with the $SU(3)_{\hat{q}}$, $SU(3)_{\hat{e}^c}$ and $SU(4)_{\hat{L}}$ epsilon tensors are understood. 
Actually, it turns out that the tensor structure forces the invariant to span over RPV couplings and light generation Yukawas, 
thus providing an automatic suppression of $\lambda''$.
Remarkably we are able to satisfy the bounds from proton decay without invoking any \emph{ad hoc} conservation or small breaking of the $U(1)_B$ symmetry, but just requiring our minimality condition on the identification of the flavor spurions.

\subsection{A toy model}
\label{toy}

When all the RPV spurions are switched on there is an overabundance of free parameters, which cannot all be fixed by the constraints from 
the neutrino sector. 
Hence we are going to consider scenarios in which only a minimal number of spurions 
are switched on in order to reproduce neutrino masses and mixings.

Our goal is to show how to link the RPV spurions with the neutrino observables, by means of a 
simplified model of neutrino masses featuring only 
the RPV couplings $\mu^i$ and $\lambda'^{i33}$. 

Notice that we work in the basis $\vev{\tilde{\nu}_i} = 0$.  
We have checked explicitly that, 
within these set of RPV couplings, 
this condition can be consistently achieved with the MFV expansion in~\eq{MFVexpSU3lang}.

The neutrino mass matrix can be split in a tree level and a one-loop 
term
\be
m_\nu = m_\nu^{\rm{(tree)}} + m_\nu^{\rm{(loop)}} \, ,
\ee
whose diagonalization through the PMNS matrix $\hat{U}$ yields
\be
m_\nu = \hat{U} \hat{m}_{\nu} \hat{U}^T \, ,
\ee
where $\hat{m}_\nu$ is the diagonal neutrino mass matrix. 
Then, assuming the dominance of the tree level contribution 
and requiring an orthogonality condition between the vectors $\mu^i$ and $\lambda'^{i33}$ 
(cf.~\app{RPVandNuMasses} for more details), the neutrino sector observables are fitted in the following analytical way:
\begin{itemize}
\item Tree level contribution
\begin{equation}
\label{eq:neutreet}
(m_{\nu}^{(\rm{tree})})^{ij} 
=  m_3 \, \hat{U}^{i3} \hat{U}^{j3} \, ,
\end{equation}
where $m_3 \approx \sqrt{\Delta m^2_{\textrm{atm}}} = 4.9 \cdot 10^{-2} \ \textrm{eV}$.
Taking $M_1 = M_2 \equiv \tilde{m} \gg M_Z$ in~\eq{eq:mefftree} of~\app{RPVandNuMasses}, we get 
\be
\label{eq:neutree}
\frac{\mu^{i}}{\mu} = 2.4 \cdot 10^{-5}   \left( \frac{ \tilde{m}}{1 \textrm{ TeV}} \right)^{1/2}  \left( \frac{ \tan \beta}{10} \right) \hat{U}^{i3} \, .
\ee
\item One-loop contribution
\begin{equation}
\label{eq:neuloopt}
(m_{\nu}^{(\rm{loop})})^{ij} =  m_2 \, \hat{U}^{i2} \hat{U}^{j2} \, , 
\end{equation}
where $m_2 \approx \sqrt{\Delta m^2_{\textrm{sol}}} = 8.7 \cdot 10^{-3} \ \textrm{eV}$. 
Taking $\mu = \tilde{m}$ and at the leading order in the MFV expansion with 
$c_{A_D} = c_{d^c} = c_q = 1$ (cf.~\eq{mlpMFV} in~\app{RPVandNuMasses}), we get
\be
\label{eq:neuloop}
(\lambda')^{i33} = 3.6 \cdot 10^{-5} 
 \left( \frac{ \tilde{m}}{1 \ \textrm{TeV}} \right)^{1/2}  \left( \frac{ \tan \beta}{10} \right)^{-1/2} \hat{U}^{i2} \, .
\ee
\end{itemize}

\section{Analysis of LFV processes}

\subsection{Case $SU(3)^4 \otimes SU(4)$} 

Once the relevant spurions are fixed in terms of the neutrino masses and mixings one can 
relate the MFV expansion to LFV processes. 
For our purposes it is enough to make an order of magnitude estimate of the processes induced in our MLFV setup, 
focusing just on the effects due to the non-diagonal entries in the sfermion mass 
matrices.
In this case the normalized branching fractions for the processes $\ell_i \rightarrow \ell_j \gamma$ 
can be naively estimated in the following way~\cite{Masina:2002mv,Paradisi:2005fk}:
\begin{equation}
\label{eq:sketchBR}
B_{\ell_i \rightarrow \ell_j\gamma}\equiv\frac{BR(\ell_i \rightarrow \ell_j \,  \gamma)}{BR(\ell_i \rightarrow \ell_j \nu_i \bar{\nu}_j)} \approx \frac{\alpha^3} {G_F^2}\frac{\delta_{ij}^2}{\tilde{m}^4} \tan^2\beta \, ,
\end{equation}
where, according to the 
MFV expansion, the flavour violating mass insertions $\delta_{ij}$ are expressed 
as combinations of neutrino masses and elements of the PMNS matrix. 
For instance in our toy model where only the couplings $\mu^i$ and $\lambda'^{i33}$ are switched on, $\delta_{ij}^{LL}$ reads (cf.~\eq{MFVexpSU3lang}) 
\begin{equation}
\label{expansdll}
\delta_{ij}^{LL}=\frac{\Delta_{ij}^{LL}}{\tilde{m}_\ell^2} \approx \frac{1}{c_L}\left( d^{2}_{L}  (\lambda')^{i33} \lambda'^*_{j33}
+ d^3_L \, \frac{\mu^{i} \mu^*_j}{|\mu|^2} \right) \, ,
\end{equation}
where $\Delta_{ij}^{LL}$ is the flavour violating part of $\tilde{m}_\ell^2$.
As it is evident from~\eq{expansdll}, the mass insertions scale like the square of the RPV parameters. 
Given the following estimation of the branching fractions in~\eq{eq:sketchBR}
\begin{equation}
B_{\ell_i \rightarrow \ell_j\gamma}
\approx  10^{-27}  \ {\left(\frac{\tilde{m}}{1 \mbox{TeV}}\right)}^{-4} {\left(\frac{\tan\beta}{10}\right)}^{2} \left( \frac{\lambda' }{10^{-5}} \right)^4 \, ,
\end{equation}
one concludes that it is not possible to accomplish observable rates, in view of the current experimental bounds 
showed in~\Table{tabsummarylfv}. 
\begin{table}[htbp]
\begin{center}
\begin{tabular}{|l|l|}
\hline
LFV process & Bound \\
\hline
$BR(\mu \rightarrow e\,\gamma)$ & $2.4 \times 10^{-12}$ \cite{Uchiyama:2011zz}\\
\hline
$BR(\tau \rightarrow e\,\gamma)$ & 3.3 $\times 10^{-8}$ \cite{Aubert:2009tk} \\
\hline
$BR(\tau \rightarrow \mu\,\gamma)$ & 4.5 $\times 10^{-8}$ \cite{Hayasaka:2007vc}\\
\hline
\end{tabular}
\end{center}
\caption{
Current experimental bounds on $\ell_i \rightarrow \ell_j \gamma$ processes. 
}
\label{tabsummarylfv}
\end{table}

It should be also mentioned that besides the contributions related to the MFV mass insertions
there are other ones due to the mixing between neutrinos and neutralinos (or charginos and charged leptons) 
and RPV vertices (see e.g.~\cite{Carvalho:2002bq,deCarlos:1996du}) which are relevant for the 
calculation of the LFV branching fractions. 
These latter contributions, however, remain of the same order of magnitude of those due to the mass insertions, 
thus leading to non-observable rates for the values of the RPV couplings fitting neutrino masses. 

Let us mention that rates of $\mu \rightarrow e \, \gamma$ closer to the experimental sensitivity can be obtained 
when neutrino masses are fitted by trilinears featuring first families indices, like for instance $\lambda'^{i11}$. 
In such a case the suppression due to the down-quark mass in the expression of the neutrino mass matrix (cf.~\eq{mlpMFV} in~\app{RPVandNuMasses}) 
allows for larger values of $\lambda'^{i11}$ even of $\mathcal{O}(10^{-2})$. However such a large coupling may be in conflict with other 
flavour violating observables \cite{Barbier:2004ez}. 
A complete analysis of such scenarios and a more realistic model for neutrino masses is postponed to future works. 

In the next subsection we are going to present a simple solution 
in order to achieve observable rates within the neutrino mass model considered here.

\subsection{Case $SU(3)^5 \otimes U(1)_{L} \otimes U(1)_{B}$}
\label{SU35LB}

We have previously seen that the contribution to LFV processes are generically well below the present experimental bounds.
This is due to fact that the spurions responsible for neutrino masses break simultaneously both 
the total lepton number and the non-abelian part of the flavor group. 
As it has been shown in~\cite{Cirigliano:2005ck,Gavela:2009cd}, in order to have measurable rates for the flavor changing radiative charged lepton decays, 
one has to separate the source of breaking of lepton number from that of LFV.

This leads us to consider a different scenario based on another subgroup of the original kinetic symmetry $G^{\rm{MSSM}}_{\rm{kin}}$ (cf.~\eq{kineticMSSM}). 
We assume that the symmetry that we want to formally restore is given by $SU(3)^5 \otimes U(1)_{L} \otimes U(1)_B$, 
where $L$ and $B$ are the total lepton and baryon number. 
The $U(1)_B$ factor is needed in order to properly suppress dangerous contributions to the proton decay rate 
(for the relevant bounds see for instance Ref.~\cite{Smirnov:1996bg,Bhattacharyya:1998dt}).

In this setup the R-parity violating couplings $\mu_i$, $\lambda$, $\lambda'$ and $\lambda''$ can be split in two parts, 
one responsible for the breaking of lepton and baryon number and the other for the breaking of the flavor group 
\begin{equation}
\mu^i=\varepsilon_{L}  \tilde{\mu}^i \, , \qquad 
\lambda=\varepsilon_{L}  \tilde{\lambda} \, , \qquad 
\lambda'=\varepsilon_{L} \tilde{\lambda}^{'} \, , \qquad 
\lambda''=\varepsilon_{B} \tilde{\lambda}^{''} \, .
\end{equation}
The quantum numbers of the flavor spurions under $SU(3)^5 \otimes U(1)_{L} \otimes U(1)_{B}$ are given by
\bea
y_U & \sim & (\bar{3},\bar{3},1,1,1)_{(0,0)}  \\
y_D & \sim & (\bar{3},1,\bar{3},1,1)_{(0,0)} \\
y_E & \sim & (1,1,1,\bar{3},\bar{3})_{(0,0)} \\
\tilde{\mu}^{i} & \sim & (1,1,1,1,\bar{3})_{(0,0)} \\
\tilde{\lambda} & \sim & (1,1,1,\bar{3},3)_{(0,0)} \\
\tilde{\lambda}^{'} & \sim & (\bar{3},1,\bar{3},1,\bar{3})_{(0,0)} \\
\tilde{\lambda}^{''} & \sim & (1,\bar{3},3,1,1)_{(0,0)} \\
\varepsilon_L & \sim & (1,1,1,1,1)_{(-1,0)} \\
\varepsilon_B & \sim & (1,1,1,1,1)_{(0,+1)}
\eea 
where the subscripts label the abelian quantum numbers.
The corresponding MFV expansion of the soft terms is reported in~\eq{MFVexpSU3BL} of~\app{gtSU35}. 

In this case the rates of the LFV processes are dominated by the lepton and baryon number preserving (but flavor changing) slepton mass insertions, 
which depend only on the
parameters $\tilde{\mu}$, $\tilde{\lambda}$, $\tilde{\lambda}^{'}$. Other RPV vertex contributions depend on quantities which violate total lepton number 
and hence are suppressed by the $\varepsilon_{L}$ factor. 

As we are going to show, peculiar correlations among physical observables will emerge due the MFV expansion. 

In the case that only the spurions $\mu^i$ and $\lambda'^{i33}$ are switched on, the relevant off-diagonal terms $i \neq j$ induced by these two spurions in $\left( \tilde{m}^2_{\ell} \right)^{i}_{j}$ 
and $a^{ij}_E$ are given by (cf.~\eq{MFVexpSU3BL} in~\app{groupth})
\be
\left( \tilde{m}^2_{\ell} \right)^{i}_{j} =  \tilde{m}^2 
\left( d^{2}_{\ell}  (\tilde{\lambda}')^{i33} \tilde{\lambda}'^*_{j33}
+ d^3_{\ell} \, \frac{\tilde{\mu}^{i} \tilde{\mu}^*_j}{\abs{\mu}^2}  \right) \, , \quad
a^{ij}_E = A \,  y_E^{jj}
\left( 
d^4_{a_E} \frac{\tilde{\mu}^*_j \tilde{\mu}^i}{\abs{\mu}^2}   + d^5_{a_E}  \tilde{\lambda}'^*_{j33} (\tilde{\lambda}')^{i33}  \right) \, .
\ee
In our framework the LL mass insertions, and thus $\left( \tilde{m}^2_{\ell} \right)^{i}_{j}$, give the dominant contribution 
to the LFV processes\footnote{The term $a_E$ is responsible for the LR mass insertions. However,  
according to the analysis of Refs.~\cite{Masina:2002mv,Paradisi:2005fk}, $\delta^{LR}$ is negligible provided that
$\delta_{ij}^{LR} \ll \left(m_{\ell_i} / \tilde{m} \right) \tan\beta \ \delta_{ij}^{LL}$.
In our case, assuming all the coefficients of the MFV expansion to be of order one, this condition translates into
$\abs{ \left( \tilde{m} / A \right)  \tan\beta } \gg 1$.
}. 
Focusing on $\delta^{LL}$, using (\ref{eq:neutreet}) and (\ref{eq:neuloopt}), we get:
\begin{multline}
\label{eq:susyinsertion}
\left( \delta^{LL} \right)^{i}_{j} = \frac{1}{ c_{\ell}} \left[ d^3_\ell \frac{\tilde{\mu}^i \tilde{\mu}^{*}_j}{\abs{\mu}^2}+d^2_\ell (\tilde{\lambda}')^{i33} \tilde{\lambda}'^*_{j33} \right] \\ 
= \frac{1}{\varepsilon_{L}^2 c_{\ell}} 
\left[ d^3_\ell {\left(\frac{ \tan\beta}{M_Z}\right)}^2 \left(\frac{M_1 M_2}{M_1 c_W^2 +M_2 s_W^2} - \frac{M_Z^2}{\mu} \sin 2 \beta \right) 
 \sqrt{\Delta m^2_{\textrm{atm}}} \, \hat{U}^{i3} (\hat{U}^{j3})^* \right. \\
\left. +d^2_\ell \frac{8 \pi^2 \tilde{m}^2}{3 \mu \tan\beta \, m_b^2} \, \sqrt{\Delta m^2_{\textrm{sol}}} \, \hat{U}^{i2} (\hat{U}^{j2})^* \right] \,.
\end{multline}
Notice that the factor $1/\varepsilon_L^2$ in the mass insertions implies an enhancement of $1/\varepsilon_L^4$ in the rates. 
Indeed it is possible to estimate the normalized branching ratios in the following way
\begin{equation}
B_{\ell_i \rightarrow \ell_j\gamma} 
\approx 10^{-27} \left( \frac{1}{\varepsilon_L} \right)^4 {\left(\frac{\tilde{m}}{1 \ \mbox{TeV}}\right)}^{-4} {\left(\frac{\tan\beta}{10}\right)}^{2} {\left(\frac{\lambda'}{10^{-5}}\right)}^4  \, ,
\end{equation}
for values of $\varepsilon_L \sim 10^{- (3 \div 4)}$ the rates of the three relevant processes 
can get close to the experimental sensitivities, depending on the values of SUSY parameters. 

Notice that the coupling $\varepsilon_L$ cannot be arbitrarily small. Indeed, by imposing the relations 
in~\eqs{eq:neutree}{eq:neuloop} 
and by requiring that the flavor violating parameters $\tilde{\mu}$ and $\tilde{\lambda}'$ 
are at most of order one, we can estimate the lower bound $\varepsilon_L \gtrsim 10^{-5}$.  

Given the potential detectability of these processes it is now interesting to compute the ratios among the normalized branching fractions of the LFV channels. 
Considering~\eq{eq:sketchBR} and~\eq{eq:susyinsertion} 
we can parametrize the ratios among the normalized LFV branching fractions as 
\begin{equation}
\label{ratiobf}
\frac{B_{\ell_j \rightarrow \ell_i\gamma}}{B_{\ell_k \rightarrow \ell_m \gamma}}=\frac{|\hat{U}^{i2} (\hat{U}^{j2})^* +c \, \hat{U}^{i3} (\hat{U}^{j3})^*|^2}{|\hat{U}^{m2} (\hat{U}^{k2})^* +c \, \hat{U}^{m3} (\hat{U}^{k3})^*|^2} \, ,
\end{equation}
where the constant $c$ is given by
\begin{equation}
\label{eq:cest}
c \approx 1.4 \times 10^{-1} \left(\frac{d_\ell^3}{d_\ell^2}\right) {\left(\frac{\tan\beta}{10}\right)}^{3} {\left(\frac{\mu}{1 \ \mbox{TeV}}\right)}{\left(\frac{\tilde{m}}{1 \ \mbox{TeV}}\right)}^{-2}{\left(\frac{M_G}{300 \ \mbox{GeV}}\right)} \, ,
\end{equation}   
with $M_1 = M_2 \equiv M_G$.
From~\eq{eq:cest} it is evident that, depending on the SUSY parameters, the mass insertions are dominated either by the trilinear ($c \ll 1$) or 
the bilinear ($c \gg 1$) couplings. 
It is possible then to identify two asymptotic regimes in which these ratios
have a simple analytical expression:
\begin{itemize}
\item $\abs{c} \ll 1$
 
In this case the mass insertions are dominated by the contribution from the trilinear couplings $(\tilde{\lambda}')^{i33}$.
Since the ratios $B_{\mu \rightarrow e\gamma}/B_{\tau \rightarrow \mu \gamma}$ and $B_{\mu \rightarrow e\gamma}/B_{\tau \rightarrow e \gamma}$ 
show only a slight dependence from the Maiorana phase $\delta$, we take $\delta=0,\pi$ and obtain 
\begin{equation}
\frac{B_{\mu \rightarrow e\gamma}}{B_{\tau \rightarrow \mu \gamma}}=
\frac{|\hat{U}^{12}|^2}{|\hat{U}^{32}|^2} = \frac{s_{12}^2 c_{13}^2}{(\mp c_{23} s_{12} s_{13} - c_{12}s_{23})^2} \approx 0.53 \div 1.75 \, ,
\end{equation}
\begin{equation}
\frac{B_{\mu \rightarrow e\gamma}}{B_{\tau \rightarrow e \gamma}} = 
\frac{|\hat{U}^{22}|^2}{|\hat{U}^{32}|^2} = \frac{(c_{12}c_{23} \mp s_{12} s_{13} s_{23})^2}{(\mp c_{23} s_{12} s_{13} - c_{12}s_{23})^2} \approx 0.37 \div 2.4 \, ,
\end{equation}   
where the extrema of the range are obtained by scanning over the 2-$\sigma$ values of the mixing angles  (cf.~\Table{tab:summaryneu}). In this case, the three normalized branching ratios are of the same order of magnitude. Notice that the LFV effects depend on the PMNS matrix $\hat{U}^{i2}$, differently with respect to other MLFV 
setups (cf.~for instance~\Table{tab:summary}).

\item $\abs{c} \gg 1$

In this case the mass insertions are dominated by the bilinear couplings $\mu^i$. 
Then we can derive the following functional behaviors for the two relevant ratios
\begin{equation}
\frac{B_{\mu \rightarrow e\gamma}}{B_{\tau \rightarrow \mu \gamma}}=  \frac{|\hat{U}^{13}|^2}{|\hat{U}^{33}|^2}= \frac{s_{13}^2}{c_{13}^2c_{23}^2}
\approx 0.007 \div 0.07 \, ,
\end{equation}
\begin{equation}
\frac{B_{\mu \rightarrow e\gamma}}{B_{\tau \rightarrow e \gamma}} = \frac{|\hat{U}^{23}|^2}{|\hat{U}^{33}|^2}  =\frac{s^2_{23}}{c_{23}^2} \approx 0.7 \div 1.6 \, .
\end{equation}
Compared to the previous case we observe an enhancement of $B_{\tau \rightarrow \mu\gamma}$
with respect to $B_{\tau \rightarrow e\gamma}$ and $B_{\mu \rightarrow e\gamma}$ . 
This results coincides with the one found in Ref.~\cite{Alonso:2011jd} in the case of inverted hierarchy of neutrino masses.

\end{itemize}

\begin{table}[htbp]
\begin{center}
\begin{tabular}{|c|c|c|}
\hline
Observable & Best fit & 2-$\sigma$ \\
\hline
$\Delta m_{\rm atm}^2$ & $2.50 \times 10^{-3} \ {\mbox{eV}}^2$ & $(2.25-2.68) \times 10^{-3} \ {\mbox{eV}}^2$ \\
\hline
$\Delta m_{\rm sol}^2$ & $7.59 \times 10^{-5} \ {\mbox{eV}}^2$ & $(7.24-7.99) \times 10^{-5} \ {\mbox{eV}}^2$  \\
\hline
$\sin^2 \theta_{12}$ & $0.312$ & $0.28-0.35$ \\
\hline
$\sin^2 \theta_{23}$ & $0.52$ & $0.41-0.61$ \\
\hline
$\sin^2 \theta_{13}$ & $0.013$ &  $0.004-0.028$ \\
\hline
\end{tabular}
\end{center}
\caption{Experimental values of the neutrino sector observables as reported in Ref.~\cite{Schwetz:2011zk}. For the PMNS matrix we have considered the PDG parametrization  \cite{Nakamura:2010zzi}. The Dirac phase $\delta$ varies in the range $[0,2\pi]$. }
\label{tab:summaryneu}
\end{table}

\begin{figure}[htbp]
\centering
\subfloat{\includegraphics[width=8 cm, height= 6 cm, angle=360]{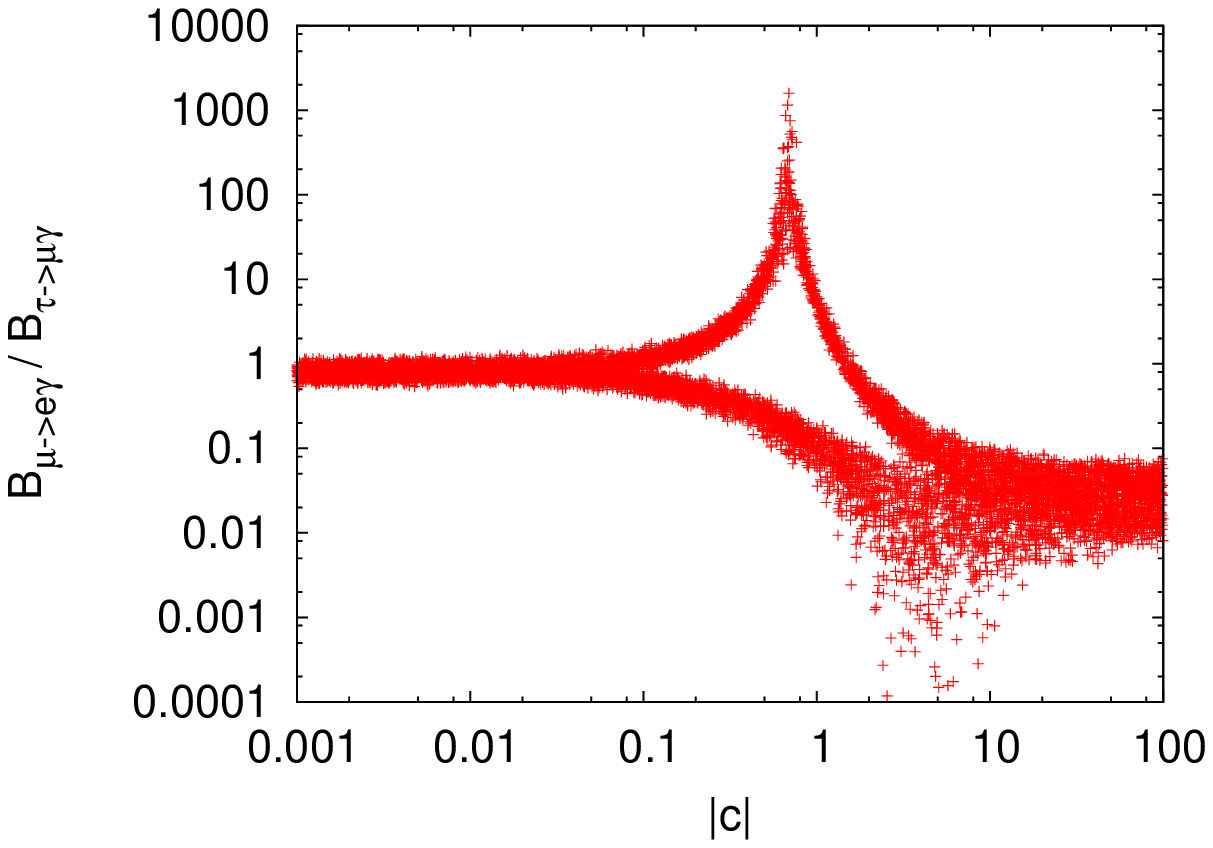}} 
\subfloat{\includegraphics[width=8 cm, height= 6 cm, angle=360]{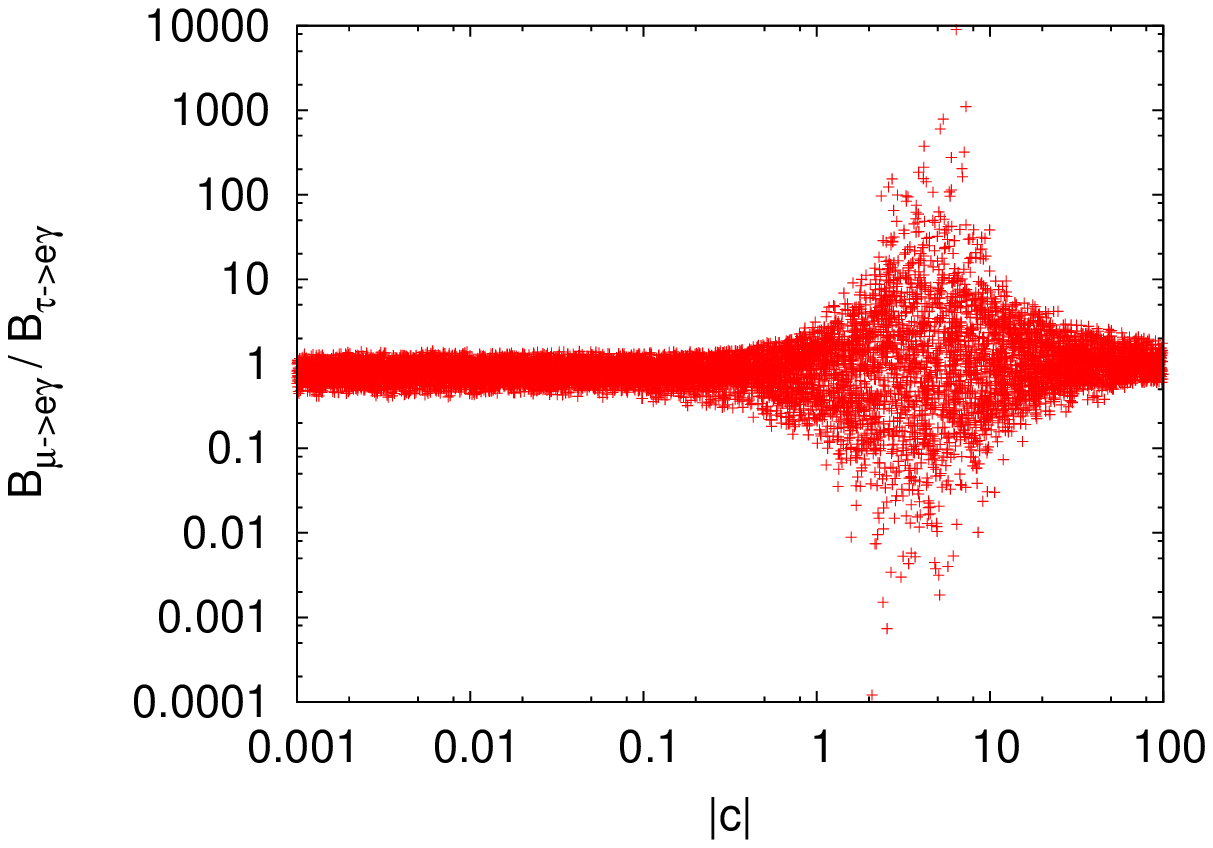}}
\caption{Ratios between normalized branching fractions of as function of $|c|$.}
\label{fig:ben2s13}
\end{figure}

\begin{figure}[htbp]
\centering
\includegraphics[width=10cm,height=8cm,angle=360]{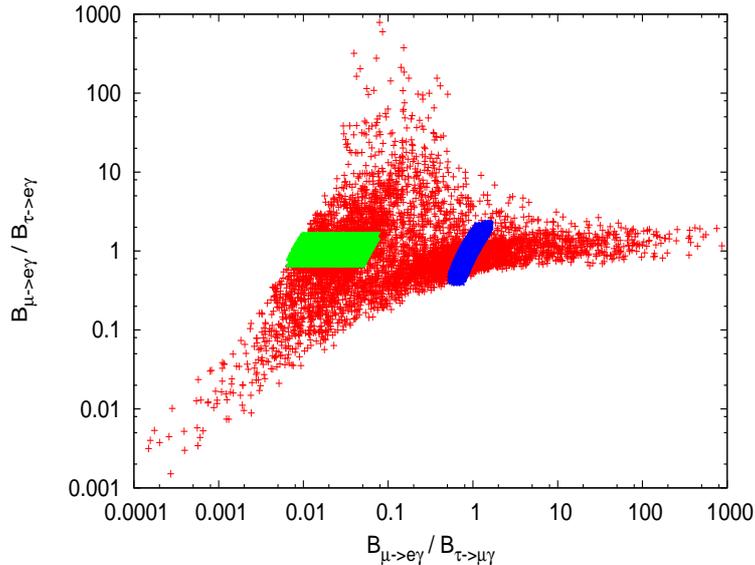}
\caption{$B_{\mu \rightarrow e\gamma}/B_{\tau \rightarrow \mu\gamma}$ versus $B_{\mu \rightarrow e\gamma}/B_{\tau \rightarrow e\gamma}$. 
The parameter $c$ is varied in the range $[-100, 100]$. 
The blue region is characterized by $\abs{c} = 0$ while the green one by $\abs{c} \gg 1$.}
\label{fig:ben2delta}
\end{figure}

Furthermore, in order to study the general case we vary the parameter $c$ in the range $[-100,100]$ and the parameters of the neutrino sector 
according to~\Table{tab:summaryneu} as before. 
The results are plotted in~\fig{fig:ben2s13}. In~\fig{fig:ben2delta} we report the correlation between the two ratios 
$B_{\mu \rightarrow e\gamma}/B_{\tau \rightarrow \mu\gamma}$ and $B_{\mu \rightarrow e\gamma}/B_{\tau \rightarrow e\gamma}$.

From~\fig{fig:ben2s13} we see that, away from the two asymptotic regimes, 
there are regions of strong enhancement/suppression of the ratios. 
Indeed the values of the $c$ parameter for which $B_{\ell_i \rightarrow \ell_j\gamma} \rightarrow 0$ are:
\begin{align}
& B_{\mu \rightarrow e\gamma} \to 0 \quad \longrightarrow \quad c =\mp \frac{c_{12}s_{23}s_{12}}{s_{23}s_{13}}
\approx \mp (2.2 \div 9.4) \qquad (\delta=0,\pi) \, , \\
& B_{\tau \rightarrow e\gamma} \rightarrow 0 \quad \longrightarrow \quad c = \pm \frac{c_{12}s_{23}s_{12}}{s_{23}s_{13}} \approx \pm (2.2 \div 9.4) 
\qquad (\delta=0,\pi) \, , \\
& B_{\tau \rightarrow \mu\gamma} \rightarrow 0 \quad \longrightarrow \quad c = \frac{c_{12}^2}{c_{13}^2} \approx (0.66 \div 0.72) \, .
\end{align}
Finally, for the purpose of comparison, we show in~\Table{tab:summary} 
the LFV parameters predicted by various MLFV models. 

\begin{table}[htbp]
\begin{center}
\begin{tabular}{|c|c|}
\hline
Model & Flavor violating parameter \\
\hline
Minimal field content \cite{Cirigliano:2005ck} & $\propto (\hat{U} \hat{m}_{\nu}^2\hat{U}^{\dagger})^{ij}$ \\
\hline 
Extended field content + CP limit \cite{Cirigliano:2005ck}& $\propto (\hat{U}\hat{m}_{\nu}\hat{U}^{\dagger})^{ij}$ \\
\hline
Extended field content + leptogenesis \cite{Cirigliano:2006nu,Branco:2006hz}  & $\propto (\hat{U}\hat{m}_{\nu}^{1/2}\,H^2\,\hat{m}_{\nu}^{1/2}\hat{U}^{\dagger})^{ij}$ \\
 \hline
$SU(3)_{\ell} \otimes SU(3)_N \to SU(3)_{\ell+N} $ \cite{Alonso:2011jd} & $\propto (\hat{U}\frac{1}{\hat{m}^2_{\nu}}\hat{U}^{\dagger})^{ij}$ \\
 \hline
MSSM wihtout R-parity (toy model) &  $\propto( \hat{U}^{i2}\hat{U}^{*j2}+c\,\hat{U}^{i3}\hat{U}^{*j3}) $\\
\hline
\end{tabular}
\end{center}
\caption{Comparative summary of MLFV models.}
\label{tab:summary}
\end{table}
We conclude commenting on other LFV processes like $\mu \rightarrow e$ conversions 
and $\ell_i \rightarrow \ell_j \ell_k \ell_k$ decays, not considered until now.
In our case these processes are determined by $\gamma$-penguin type diagrams~\cite{Arganda:2005ji,Brignole:2004ah} 
and turn out to have the same flavor structure of $\ell_i \rightarrow \ell_j \gamma$. 
This implies in particular that the decays in three leptons have similar patterns of enhancement/suppression of those discussed above. 
Notice however that the radiative decays are the processes most severely constrained by the experiments.

\section{Conclusions}

In this work we have presented the general supersymmetric version of MFV including also 
the RPV terms as the irreducible sources of the flavour symmetry breaking. 
If the RPV couplings are responsible for neutrino masses, the framework can be also viewed as an 
extension of MFV to the lepton sector.   

An important aspect stressed throughout the paper is that the global symmetry of the kinetic term of the MSSM lagrangian
is enhanced with respect to that of the SM. Indeed the superfields $\hat{\ell}_i$ and $\hat{h}_d$ 
can be rearranged in a 4-dimensional flavour multiplet $\hat{L}$, whose kinetic term is invariant 
under $U(4)_{\hat{L}}$ unitary transformations. 
This gives us the possibility to consider as the most general flavour symmetry the non-abelian group $SU(3)^4 \otimes SU(4)_{\hat{L}}$. 
In such a case the breaking of the total lepton number and that of lepton flavour number are linked together, 
thus generically implying small effects in LFV physics. 

On the other hand the separation between the breaking of lepton number and lepton flavour number leads to an interesting phenomenology. 
This is the motivation to consider our second scenario based on the $SU(3)^5 \otimes U(1)_L \otimes U(1)_B$ flavour symmetry.  
This last option yields peculiar correlations among the branching ratios of the $\ell_i \rightarrow \ell_j \gamma$ 
processes. 

Several interesting possibilities could be considered for future investigations both from a theoretical 
and a phenomenological point of view.

\section*{Acknoledgments}
We thank Federica Bazzocchi, Ben Grinstein, Jernej  Kamenik and Federico Mescia for useful discussions 
and Gino Isidori for encouragement. 
We also thank Andrea Romanino and Marco Serone for further useful comments.
This work was supported in part by the National Science Foundation under Grant No.~1066293 and the hospitality of the Aspen Center for Physics.
The work of LDL was supported by the DFG through the SFB/TR 9 ``Computational Particle Physics''; 
he would also like to thank the CP3-Origins group at the University of Southern Denmark for the warm 
hospitality and the partial support during the final stages of this work.

\appendix

\section{Notation}
\label{appnotation}
In this Appendix we define both the $SU(3)^4 \otimes SU(4)$ and $SU(3)^5$ notations and provide 
the translation between the two languages.
\begin{itemize}
\item $SU(3)^4 \otimes SU(4)$ notation
\begin{equation}
\label{SU(4)not}
\begin{array}{rc}
W=& Y_U^{ij} \hat{q}_i \hat{u}^c_j \hat{h}_u + Y_D^{\alpha ij} \hat{L}_{\alpha} \hat{q}_i \hat{d}^c_j 
+ \frac{1}{2} Y_E^{\alpha \beta i} \hat{L}_{\alpha} \hat{L}_{\beta} \hat{e}^c_i + \mu^{\alpha} \hat{h}_u \hat{L}_{\alpha}  
+  \frac{1}{2}  (\lambda^{''} )^{ijk}  \hat{u}^c_i   \hat{d}^c_j  \hat{d}^c_k  \\ \\
- \mathcal{L}_{\textrm{soft}}=& \tfrac{1}{2} ( M_1 \, \tilde{B} \tilde{B} + M_2 \, \tilde{W} \tilde{W} + M_3 \, \tilde{g} \tilde{g} + \text{h.c.} ) \\
+& \sum_{F} \tilde{F}^{\dagger} \tilde{m}^2_F \tilde{F} + \tilde{m}^2_{h_u} h^{\ast}_u h_u + ( B^{\alpha} h_u \tilde{L}_{\alpha} + \textrm{h.c.} ) \\
+& A_U^{ij} \tilde{q}_i \tilde{u}^c_j h_u + A_D^{\alpha ij} \tilde{L}_{\alpha} \tilde{q}_i \tilde{d}^c_j 
+ \frac{1}{2} A_E^{\alpha \beta j} \tilde{L}_{\alpha} \tilde{L}_{\beta} \tilde{e}^c_j 
+ \frac{1}{2}  (A_{\lambda^{''}} )^{ijk}  \tilde{u}^c_i   \tilde{d}^c_j  \tilde{d}^c_k + \textrm{h.c.} \\
\end{array}
\end{equation}
with $F= \{ q,u^c,d^c,e^c,L \}$.
\item $SU(3)^5$ notation
\begin{equation}
\label{SU(3)not}
\begin{array}{rc}
W_{\textrm{RPC}}=& y_U^{ij} \hat{q}_i \hat{u}^c_j \hat{h}_u + y_D^{ij} \hat{h}_d \hat{q}_i \hat{d}^c_j +  y_E^{ij} \hat{h}_d \hat{\ell}_i \hat{e}^c_j 
+ \mu \, \hat{h}_u \hat{h}_d  \\ \\
W_{\textrm{RPV}}=& \mu^{i} \hat{h}_u \hat{\ell}_i + \frac{1}{2} \lambda^{ijk} \hat{\ell}_i \hat{\ell}_j \hat{e}^c_k 
+ (\lambda^{'} )^{ijk}  \hat{\ell}_i \hat{q}_j  \hat{d}^c_k +  \frac{1}{2}  (\lambda^{''} )^{ijk}  \hat{u}^c_i   \hat{d}^c_j  \hat{d}^c_k  \\ \\
- \mathcal{L}_{\textrm{soft}}^{\textrm{RPC}}=& \tfrac{1}{2} ( M_1 \, \tilde{B} \tilde{B} + M_2 \, \tilde{W} \tilde{W} + M_3 \, \tilde{g} \tilde{g} + \text{h.c.} ) \\
+& \sum_{f} \tilde{f}^{\dagger} \tilde{m}^2_f \tilde{f} + \tilde{m}^2_{h_u} h^{\ast}_u h_u +  \tilde{m}^2_{h_d} h^{\ast}_d h_d 
+ \left( b \, h_u h_d + \textrm{h.c.} \right) \\
+& a_U^{ij} \tilde{q}_i \tilde{u}^c_j h_u + a_D^{ij} h_d \tilde{q}_i \tilde{d}^c_j + a_E^{ij} h_d \tilde{\ell}_i \tilde{e}^c_j + \textrm{h.c.}  \\ \\
- \mathcal{L}_{\textrm{soft}}^{\textrm{RPV}}=& (\tilde{m}^2_d)^{i} h_d^{\ast} \tilde{\ell}_i + b^i h_u \tilde{\ell}_i + \textrm{h.c.}  \\
+& \frac{1}{2} (a_{\lambda})^{ijk} \tilde{\ell}_i   \tilde{\ell}_j  \tilde{e}^c_k + (a_{\lambda^{'} })^{ijk}  \tilde{\ell}_i \tilde{q}_j  \tilde{d}^c_k 
+ \frac{1}{2}  (a_{\lambda^{''} })^{ijk}  \tilde{u}^c_i   \tilde{d}^c_j  \tilde{d}^c_k  + \textrm{h.c.} \\
\end{array}
\end{equation}
with $f= \{ q,u^c,d^c,e^c,\ell \}$.
\end{itemize}
Let us define 
\begin{equation}
\hat{L}_{i} \equiv \hat{\ell}_i \, , \qquad \hat{L}_{4} \equiv \hat{h}_d \, , \qquad \tilde{L}_4 \equiv h_d \, ,
\end{equation}
then, by comparing~\eq{SU(3)not} with~\eq{SU(4)not}, the following identifications follow
\begin{equation}
\label{dictionary}
\begin{array}{cccccccccccc}
& y_U^{ij} = Y_U^{ij} & \quad
& y_D^{ij} = Y_D^{4ij} & \quad
& y_E^{ij} = Y_E^{4ij} & \quad
& \mu = \mu^4 & \\
& \lambda^{ijk} = Y_E^{ijk} & \quad
& \left( \lambda '  \right)^{ijk} = Y_D^{ijk} & \quad
& \left( \tilde{m}_\ell \right) ^{i}_j = \left( \tilde{m}_L \right) ^{i}_j & \quad
& \tilde{m}_{h_d} = \left( \tilde{m}_L \right) ^4_4 & \\ 
& \left( \tilde{m}_d \right) ^{i} =  \left( \tilde{m}_L \right) ^{i}_4 & \quad
& b  =  B^4 & \quad
& b^i  =  B^i & \quad
& a_U^{ij} = A_U^{ij} & \\
& a_D^{ij} = A_D^{4ij} & \quad
& a_E^{ij} = A_E^{4ij} & \quad
& (a_{\lambda})^{ijk} = A_E^{ijk} & \quad
& (a_{\lambda^{'} })^{ijk} = A_D^{ijk} & \, .
\end{array}
\end{equation}

\section{MFV expansion}
\label{groupth}

\subsection{$SU(3)^4 \otimes SU(4)$}

Soft terms:
\begin{equation}
\begin{array}{rcl}
\tilde{m}^2_q & \sim & (8,1,1,1,1) \oplus (1,1,1,1,1)  \\
\tilde{m}^2_{u^c} & \sim & (1,8,1,1,1) \oplus (1,1,1,1,1) \\
\tilde{m}^2_{d^c}  & \sim & (1,1,8,1,1) \oplus (1,1,1,1,1)  \\
\tilde{m}^2_{e^c}  & \sim & (1,1,1,8,1) \oplus (1,1,1,1,1)  \\
\tilde{m}^2_{L}  & \sim & (1,1,1,1,15) \oplus (1,1,1,1,1)  \\
B  & \sim & (1,1,1,1,\bar{4}) \\
A_U & \sim & (\bar{3},\bar{3},1,1,1)  \\
A_D & \sim & (\bar{3},1,\bar{3},1,\bar{4}) \\
A_E & \sim & (1,1,1,\bar{3},6) \\
\end{array}
\end{equation}
The MFV expansion of the soft terms in both the $SU(3)^4 \otimes SU(4)$ and the $SU(3)^5$ languages 
is provided respectively in~\eq{MFVexpSU4} and~\eq{MFVexpSU3lang} of~\sect{flavgroupMSSM}.

\subsection{$SU(3)^5 \otimes U(1)_L \otimes U(1)_B$}
\label{gtSU35}

Soft terms:
\begin{equation}
\begin{array}{rcl}
\tilde{m}^2_q & \sim & (8,1,1,1,1)_{(0,0)} \oplus (1,1,1,1,1)_{(0,0)} \\
\tilde{m}^2_{u^c} & \sim & (1,8,1,1,1)_{(0,0)} \oplus (1,1,1,1,1)_{(0,0)} \\
\tilde{m}^2_{d^c}  & \sim & (1,1,8,1,1)_{(0,0)}  \oplus (1,1,1,1,1)_{(0,0)} \\
\tilde{m}^2_{e^c}  & \sim & (1,1,1,8,1)_{(0,0)} \oplus (1,1,1,1,1)_{(0,0)}  \\
\tilde{m}^2_{l} & \sim & (1,1,1,1,8)_{(0,0)} \oplus (1,1,1,1,1)_{(0,0)} \\
\left( \tilde{m}^2_d \right)^i & \sim & (1,1,1,1,\bar{3})_{(+1,0)} \\  
b^i & \sim & (1,1,1,1,\bar{3})_{(+1,0)}\\
a_U & \sim & (\bar{3},\bar{3},1,1,1)_{(0,0)}  \\
a_D & \sim & (\bar{3},1,\bar{3},1,1)_{(0,0)} \\
a_E & \sim & (1,1,1,\bar{3},\bar{3})_{(0,0)} \\
a_{\lambda} & \sim & (1,1,1,\bar{3},3)_{(+1,0)} \\
a_{\lambda^{'}} & \sim & (\bar{3},1,\bar{3},1,\bar{3})_{(+1,0)} \\
a_{\lambda^{''}} & \sim & (1,\bar{3},3,1,1)_{(0,-1)}
\end{array}
\end{equation}

The MFV expansion reads
\begin{equation}
\label{MFVexpSU3BL}
\begin{array}{rcl}

\left( \tilde{m}^2_q \right)^{i}_j & = & \tilde{m}^2 
\left( c_q \delta^i_j + d^{1}_{q} (y_U y_U^{\dagger})^i_j + d^{(21)}_{q} (y_D y_D^{\dagger})^i_j + d^{(22)}_{q} (\tilde{\lambda}')^{lik}   \tilde{\lambda}'^*_{ljk} \right) \\

\left( \tilde{m}^2_{u^c} \right)^{i}_j & = & \tilde{m}^2 
\left( c_{u^c} \delta^i_j +  d^{1}_{u^c}  (y_U^{\dagger}  y_U )^i_j 
+ d^{2}_{u^c} (\tilde{\lambda}^{''} )^{ikl} (\tilde{\lambda}^{''\ast} )_{jkl} \right)  \\

\left( \tilde{m}^2_{d^c} \right)^{i}_j & = & \tilde{m}^2 
\left( c_{d^c} \delta^i_j +  d^{(11)}_{d^c} (y_D^{\dagger} y_D)^i_j + d^{(12)}_{d^c} (\tilde{\lambda}')^{lki}   \tilde{\lambda}'^*_{lkj} 
+ d^{2}_{d^c} (\tilde{\lambda}^{''} )^{kil} (\tilde{\lambda}^{''\ast} )_{kjl} \right)   \\

\left( \tilde{m}^2_{e^c} \right)^{i}_j & = & \tilde{m}^2 
\left( c_{e^c} \delta^i_j +  d^{(11)}_{e^c} (y_E^{\dagger}  y_E)^i_j +  d^{(12)}_{e^c}  \tilde{\lambda}^{lki} \tilde{\lambda}^*_{lkj} \right)  \\

\left( \tilde{m}^2_{\ell} \right)^{i}_{j} & = & \tilde{m}^2 
\left( c_{\ell}  \delta^{i}_{j} + d^{(11)}_{\ell} ( y_{E} y_E^{\dagger})^i_j + d^{(12)}_{\ell} \tilde{\lambda}^{ilk} \tilde{\lambda}^*_{jlk}  + d^{2}_{\ell}  (\tilde{\lambda}')^{ilk} \tilde{\lambda}'^*_{jlk} + d^3_{\ell} \, \tilde{\mu}^{i} \tilde{\mu}^*_j  / |\mu|  \right) \\

\left( \tilde{m}^2_{d} \right)^{i} & = &\tilde{m}^2 \ \varepsilon_L  \left(  d^1_d \, \tilde{\mu}^i  / |\mu|
+ d^2_d \, (\tilde{\lambda}')^{ilk} (y^*_D)_{lk} \,  + d^3_d \, (\tilde{\lambda}')^{ilk} \tilde{\lambda}'^*_{plk} \, \tilde{\mu}^p  / |\mu| \right. \\ 
& & \left. +
d^4_d \, (\tilde{\lambda})^{ilk} (y^*_E)_{lk} \, + d^5_d \, (\tilde{\lambda})^{ilk} \tilde{\lambda}^*_{plk} \, \tilde{\mu}^p  / |\mu| +
d^6_d \, (y_E y_E^{\dagger})^i_p \, \tilde{\mu}^p  / |\mu| \right)\\

b^i &=&  \tilde{m}^2 \ \varepsilon_L  \left(  c_{b} \, \tilde{\mu}^i  / |\mu|
+ d^{(11)}_b (\tilde{\lambda}')^{ilk} (y^*_D)_{lk} + d^{(12)}_b (\tilde{\lambda}')^{ilk} \tilde{\lambda}'^*_{plk} \, \tilde{\mu}^p  / |\mu| \right. \\ 
& & \left. +
d^{(21)}_b  (\tilde{\lambda})^{ilk} (y^*_E)_{lk} + d^{(22)}_b (\tilde{\lambda})^{ilk} \tilde{\lambda}^*_{plk} \, \tilde{\mu}^p / |\mu|
+d^{(23)}_b  (y_E y_E^{\dagger})^i_p \, \tilde{\mu}^p / |\mu| \right)\\
 
a^{ij}_U &=& A \left( c_{a_U} y^{ij}_U + \dots  \right) \\

a^{ij}_D &=& A \left( c_{a_D} y^{ij}_D + \dots  \right) \\

a^{ij}_E &=& A 
\left( 
c_{a_E} y^{ij}_E  
+ d^1_{a_E} (y_E y_E^{\dagger} y_E)^{ij} 
+ d^2_{a_E} \tilde{\lambda}^{ikj} \tilde{\lambda}'^*_{klm} y_D^{lm} +
d^3_{a_E} y_E^{kj} \tilde{\mu}^*_k \tilde{\mu}^i  / |\mu| \right. \\ 
& & 
+ d^4_{a_E} y_E^{kj} \tilde{\lambda}'^*_{klm} (\tilde{\lambda}')^{ilm} 
+d^5_{a_E} y_E^{im} \tilde{\lambda}^*_{klm} \tilde{\lambda}^{klj}
+d^6_{a_E} y_E^{kj} \tilde{\lambda}^*_{klm} \tilde{\lambda}^{ilm} \\ 
& & 
+ d^7_{a_E} y_E^{km} \tilde{\lambda}^*_{klm} \tilde{\lambda}^{lij}
+ d^8_{a_E} \epsilon_{klm} \tilde{\lambda}^{ikj} \tilde{\mu}^l \tilde{\mu}^m + d^{9}_{a_E} \epsilon^{ikl} \epsilon^{jmn} (y_E^\ast)_{kl} (y_E^\ast)_{mn}  \\
& &
+ d^{10}_{a_E} \epsilon^{ikl} \epsilon^{jmn} \tilde{\mu}^p (y_E^\ast)_{pm} \tilde{\lambda}^*_{kln}
+ d^{11}_{a_E} \epsilon^{ikl} \epsilon^{jmn} \tilde{\mu}^p (y_E^\ast)_{km} \tilde{\lambda}^*_{lpn}  \\
& & \left. 
+ d^{12}_{a_E} \tilde{\lambda}^{ikj} \tilde{\mu}^*_k  / |\mu| \right) \\

(a_{\lambda})^{ijk} &=& A \ \varepsilon_L \left( c_{A_E} \tilde{\lambda}^{ijk} + \dots  \right) \\

(a_{\lambda^{'}})^{ijk}&=& A \ \varepsilon_L  \left( c_{A_D} (\tilde{\lambda}')^{ ijk} + \dots  \right) \\

a_{\lambda^{''}}^{ijk} &=& A \ \varepsilon_B  \left( c_{A_{\lambda^{''}}}  (\tilde{\lambda}^{''})^{ijk} + \dots  \right)  \, ,
\end{array}
\end{equation}
up to two flavor spurions and only one in $\varepsilon_L$ or $\varepsilon_B$. 
Just in the case of $a_E$ we consider the expansion up to three spurions.

\section{Review of RPV contributions to neutrino masses}
\label{RPVandNuMasses}

The neutrino mass matrix receives contributions both at the tree level and from loops~\cite{Barbier:2004ez}. 
We briefly review for convenience here the general formulae.  

In the basis where only $h_d$ develops an electroweak VEV, the tree level contribution to neutrino masses 
is due to the RPV mixings among neutrinos and higgsinos, proportional to the parameters $\mu_i$.
The tree level neutral fermion mass matrix in the basis $(L_{\alpha} , h_u, \tilde{B}, \tilde{W})$ reads
\begin{equation}
M_{\nu} =\left(
\begin{array}{cc}
0 & m_{RPV}\\
m^{T}_{RPV} & M_N
\end{array}
\right) \qquad 
\mbox{with} \qquad 
m_{RPV}=\left(
\begin{array}{cccc}
0 & -\mu_1 & 0 & 0\\
0 & -\mu_2 & 0 & 0\\
0 & -\mu_3 & 0 & 0
\end{array}
\right) \, ,
\end{equation}
while $M_N$ is the $4 \times 4$ neutralino mass matrix
\begin{multline}
M_N = \\
\left(
\begin{array}{cccc}
0 & -\mu & \sin\beta\sin{\theta}_W M_Z & -\sin\beta\cos{\theta}_W M_Z \\
-\mu & 0 & -\cos\beta\sin{\theta}_W M_Z & \cos\beta\cos{\theta}_W M_Z \\
\sin\beta\sin{\theta}_W M_Z & -\cos\beta\sin{\theta}_W M_Z & M_1 & 0\\
 -\sin\beta\cos{\theta}_W M_Z & \cos\beta\cos{\theta}_W M_Z & 0 & M_2 
\end{array}
\right)\,.
\end{multline}
Under the hypothesis $m_{RPV} \ll M_N$ the matrix $M_{\nu}$ can be perturbatively diagonalized,  
thus yielding for the three lightest neutrino mass matrix
\begin{equation}
\label{numatrixtree}
(m^{\rm{(tree)}}_{\nu})^{ij} \approx 
- (m_{RPV}M_N^{-1}m_{RPV}^T)^{ij} = m_{\nu}^{\rm{(tree)}}
\frac{ \mu^{i} \mu^j }{\sum_i |\mu_i^2|} \, ,
\end{equation}
where
\begin{equation}
\label{eq:mefftree}
m_{\nu}^{\rm{(tree)}} = 
\frac{\left(M_1 \cos^2 \theta_W+M_2 \sin^2 \theta_W\right) M_Z^2 \cos^2 \beta}
{\mu \left( (M_1 \cos^2\theta_W+M_2 \sin^2\theta_W) M^2_Z \sin2\beta - M_1 M_2 \, \mu \right)} \times \sum_i |\mu_i^2| \, .
\end{equation}


The tree-level neutrino mass matrix is a rank-one matrix.
In order to complete the neutrino spectrum one has to go at the loop level. 
One-loop neutrino masses get contributions from many diagrams involving different combinations of the coupling $\mu_i$, $\lambda'$ and $\lambda$. 
Under reasonable assumptions on the SUSY parameters (see e.g.~\cite{Davidson:2000ne,Chun:1999bq}),
one can focus the attention just on the contribution coming from the 
trilinear terms $\lambda$ and $\lambda'$. In the basis where the down-quark and the charged-lepton 
mass matrices are diagonal, one finds~\cite{Barbier:2004ez}
\begin{align}
(m_{\nu}^{(\lambda'\lambda')})^{ij} &= 
\frac{3}{16 \pi^2} \sum_{k,l,m}
  \lambda'^{ikl} \lambda'^{jmk}\, \hat{m}_{D_k}\
  \frac{(\tilde{m}^{d\, 2}_{\scriptscriptstyle{LR}})_{ml}}
  {m^2_{\tilde d_{Rl}} - m^2_{\tilde d_{Lm}}}\,
  \ln \left( \frac{m^2_{\tilde d_{Rl}}}{m^2_{\tilde d_{Lm}}} \right)
  +\ (i \leftrightarrow j) \, , \\
(m_{\nu}^{(\lambda\lambda)})^{ij} &= 
\frac{1}{16 \pi^2} \sum_{k,l,m}
  \lambda^{ikl} \lambda^{jmk}\, \hat{m}_{E_k}\
  \frac{(\tilde{m}^{e\, 2}_{\scriptscriptstyle{LR}})_{ml}}
  {m^2_{\tilde e_{Rl}} - m^2_{\tilde e_{Lm}}}\,
  \ln \left( \frac{m^2_{\tilde e_{Rl}}}{m^2_{\tilde e_{Lm}}} \right)
  +\ (i \leftrightarrow j) \, ,
\end{align}
which at the leading order in the MFV expansion read
\begin{eqnarray}
\label{mlpMFV}
(m_{\nu}^{(\lambda'\lambda')})^{ij} &=& \frac{3}{8 \pi^2} \frac{\tilde{m} \, c_{A_D}-\mu \tan \beta}{\tilde{m}^2}\frac{1}{c_{d^c}-c_q} \ln \left( \frac{c_{d^c}}{c_q} \right) (\lambda')^{ikl} (\lambda')^{jlk} \hat{m}_{D_{k}} \hat{m}_{D_{l}} \, , \\
\label{mlMFV}
(m_{\nu}^{(\lambda\lambda)})^{ij} &=& \frac{1}{8 \pi^2} \frac{\tilde{m} \, c_{A_E}-\mu \tan \beta}{\tilde{m}^2}\frac{1}{c_{e^c}-c_L} \ln \left( \frac{c_{e^c}}{c_L} \right) \lambda^{ikl} \lambda^{jlk} \hat{m}_{E_{k}} \hat{m}_{E_{l}} \, .
\end{eqnarray}
In this work we have considered a simplified model of neutrino masses with only the couplings 
$\vec{\mu}=(\mu^1,\mu^2,\mu^3)$ and 
$\vec{\lambda'}=(\lambda'^{133}, \lambda'^{233}, \lambda'^{333})$ set to non-zero values. 
The neutrino mass spectrum is thus obtained by the diagonalization of a matrix of the form:
\be
\label{mnutreeloop}
(m_\nu)^{ij} =c_0 \, \mu^i \mu^j
+c_1 \, \lambda'^{i33} \lambda'^{j33} \, ,
\ee
with
\be
c_0=\frac{\left(M_1 \cos^2 \theta_W+M_2 \sin^2 \theta_W\right) M_Z^2 \cos^2 \beta}
{\mu \left( (M_1 \cos^2\theta_W+M_2 \sin^2\theta_W) M^2_Z \sin2\beta - M_1 M_2 \, \mu \right)}
\ee
and
\be
c_1=\frac{3}{8 \pi^2}\frac{m^2_b \tilde{m}^{b2}_{LR}}{m^2_{\tilde{b}_R}-m^2_{\tilde{b}_L}}\ln\left(\frac{m_{\tilde{b}_R}^2}{m_{\tilde{b}_L}^2}\right) \, .
\ee
Under the assumption that the leading contribution to the atmospheric neutrino observables comes from the tree level term, 
the mass matrix in \eq{mnutreeloop} can be perturbatively diagonalized along the lines of~\cite{Hirsch:2000ef, Diaz:2003as}, obtaining at the leading order
\be
m_{{\nu}_3}\approx c_0 |\vec{\mu}|^2 \, , \qquad 
m_{{\nu}_2}\approx c_1 \frac{|\vec{\mu} \times \vec{\lambda}^{'}|^2}{|\vec{\mu}|^2} \, , \qquad
m_{{\nu}_1}=0 \, .
\ee
The orthogonality condition mentioned in~\sect{toy}, namely $|\vec{\mu} \times \vec{\lambda}^{'}|^2/|\vec{\mu}|^2=|\vec{\lambda}^{'}|^2$, 
ensures that $\lambda'^{i33}$ is completely responsible for the solar neutrino observables, 
thus allowing us to express the LFV branching fractions 
in terms of the neutrino mass observables in a simple analytical way (cf.~e.g.~\eq{ratiobf}).


\bibliography{bibfile}{}

\begin{thebibliography}{10}

\bibitem{Isidori:2010kg}
G.~Isidori, Y.~Nir, and G.~Perez, ``{Flavor Physics Constraints for Physics
  Beyond the Standard Model},'' {\em Ann.Rev.Nucl.Part.Sci.}, vol.~60, p.~355,
  2010, 1002.0900.

\bibitem{Chivukula:1987py}
R.~Chivukula and H.~Georgi, ``{Composite Technicolor Standard Model},'' {\em
  Phys.Lett.}, vol.~B188, p.~99, 1987.

\bibitem{Hall:1990ac}
L.~Hall and L.~Randall, ``{Weak scale effective supersymmetry},'' {\em
  Phys.Rev.Lett.}, vol.~65, pp.~2939--2942, 1990.

\bibitem{Buras:2000dm}
A.~Buras, P.~Gambino, M.~Gorbahn, S.~Jager, and L.~Silvestrini, ``{Universal
  unitarity triangle and physics beyond the standard model},'' {\em
  Phys.Lett.}, vol.~B500, pp.~161--167, 2001, hep-ph/0007085.

\bibitem{D'Ambrosio:2002ex}
G.~D'Ambrosio, G.~Giudice, G.~Isidori, and A.~Strumia, ``{Minimal flavor
  violation: An Effective field theory approach},'' {\em Nucl.Phys.},
  vol.~B645, pp.~155--187, 2002, hep-ph/0207036.

\bibitem{Bona:2007vi}
M.~Bona {\em et~al.}, ``{Model-independent constraints on $\Delta$ F=2
  operators and the scale of new physics},'' {\em JHEP}, vol.~0803, p.~049,
  2008, 0707.0636.

\bibitem{Hurth:2008jc}
T.~Hurth, G.~Isidori, J.~F. Kamenik, and F.~Mescia, ``{Constraints on New
  Physics in MFV models: A Model-independent analysis of $\Delta$ F = 1
  processes},'' {\em Nucl.Phys.}, vol.~B808, pp.~326--346, 2009, 0807.5039.

\bibitem{Cirigliano:2005ck}
V.~Cirigliano, B.~Grinstein, G.~Isidori, and M.~B. Wise, ``{Minimal flavor
  violation in the lepton sector},'' {\em Nucl.Phys.}, vol.~B728, pp.~121--134,
  2005, hep-ph/0507001.

\bibitem{Cirigliano:2006su}
V.~Cirigliano and B.~Grinstein, ``{Phenomenology of minimal lepton flavor
  violation},'' {\em Nucl.Phys.}, vol.~B752, pp.~18--39, 2006, hep-ph/0601111.

\bibitem{Cirigliano:2006nu}
V.~Cirigliano, G.~Isidori, and V.~Porretti, ``{CP violation and Leptogenesis in
  models with Minimal Lepton Flavour Violation},'' {\em Nucl.Phys.}, vol.~B763,
  pp.~228--246, 2007, hep-ph/0607068.

\bibitem{Davidson:2006bd}
S.~Davidson and F.~Palorini, ``{Various definitions of Minimal Flavour
  Violation for Leptons},'' {\em Phys.Lett.}, vol.~B642, pp.~72--80, 2006,
  hep-ph/0607329.

\bibitem{Branco:2006hz}
G.~C. Branco, A.~J. Buras, S.~Jager, S.~Uhlig, and A.~Weiler, ``{Another look
  at minimal lepton flavour violation, $\ell_i \rightarrow \ell_j \gamma$,
  leptogenesis, and the ratio $M_{\nu}/\Lambda_{\rm{LFV}}$},'' {\em JHEP},
  vol.~0709, p.~004, 2007, hep-ph/0609067.

\bibitem{Gavela:2009cd}
M.~Gavela, T.~Hambye, D.~Hernandez, and P.~Hernandez, ``{Minimal Flavour Seesaw
  Models},'' {\em JHEP}, vol.~0909, p.~038, 2009, 0906.1461.

\bibitem{Filipuzzi:2009xr}
A.~Filipuzzi and G.~Isidori, ``{Violations of lepton-flavour universality in $P
  \rightarrow \ell \nu$ decays: A Model-independent analysis},'' {\em
  Eur.Phys.J.}, vol.~C64, pp.~55--62, 2009, 0906.3024.

\bibitem{Alonso:2011jd}
R.~Alonso, G.~Isidori, L.~Merlo, L.~A. Munoz, and E.~Nardi, ``{Minimal flavour
  violation extensions of the seesaw},'' {\em JHEP}, vol.~1106, p.~037, 2011,
  1103.5461.

\bibitem{Barbier:2004ez}
R.~Barbier, C.~Berat, M.~Besancon, M.~Chemtob, A.~Deandrea, {\em et~al.},
  ``{R-parity violating supersymmetry},'' {\em Phys.Rept.}, vol.~420,
  pp.~1--202, 2005, hep-ph/0406039.

\bibitem{Hall:1983id}
L.~J. Hall and M.~Suzuki, ``{Explicit R-Parity Breaking in Supersymmetric
  Models},'' {\em Nucl.Phys.}, vol.~B231, p.~419, 1984.

\bibitem{Hempfling:1995wj}
R.~Hempfling, ``{Neutrino Masses and Mixing Angles in SUSY-GUT Theories with
  explicit R-Parity Breaking},'' {\em Nucl. Phys.}, vol.~B478, pp.~3--30, 1996,
  hep-ph/9511288.

\bibitem{Nilles:1996ij}
H.-P. Nilles and N.~Polonsky, ``{Supersymmetric neutrino masses, R symmetries,
  and the generalized mu problem},'' {\em Nucl. Phys.}, vol.~B484, pp.~33--62,
  1997, hep-ph/9606388.

\bibitem{Nardi:1996iy}
E.~Nardi, ``{Renormalization group induced neutrino masses in supersymmetry
  without R-parity},'' {\em Phys.Rev.}, vol.~D55, pp.~5772--5779, 1997,
  hep-ph/9610540.

\bibitem{Chun:1999bq}
E.~J. Chun and S.~K. Kang, ``{One loop corrected neutrino masses and mixing in
  supersymmetric standard model without R-parity},'' {\em Phys.Rev.}, vol.~D61,
  p.~075012, 2000, hep-ph/9909429.

\bibitem{Davidson:2000ne}
S.~Davidson and M.~Losada, ``{Basis independent neutrino masses in the R(p)
  violating MSSM},'' {\em Phys. Rev.}, vol.~D65, p.~075025, 2002,
  hep-ph/0010325.

\bibitem{Lee:1984kr}
I.-H. Lee, ``{Lepton Number Violation in Softly Broken Supersymmetry},'' {\em
  Phys. Lett.}, vol.~B138, p.~121, 1984.

\bibitem{Lee:1984tn}
I.-H. Lee, ``{Lepton Number Violation in Softly Broken Supersymmetry. 2},''
  {\em Nucl. Phys.}, vol.~B246, p.~120, 1984.

\bibitem{Dawson:1985vr}
S.~Dawson, ``{R-Parity Breaking in Supersymmetric Theories},'' {\em Nucl.
  Phys.}, vol.~B261, p.~297, 1985.

\bibitem{Banks:1995by}
T.~Banks, Y.~Grossman, E.~Nardi, and Y.~Nir, ``{Supersymmetry without R-parity
  and without lepton number},'' {\em Phys. Rev.}, vol.~D52, pp.~5319--5325,
  1995, hep-ph/9505248.

\bibitem{Hirsch:2000ef}
M.~Hirsch, M.~Diaz, W.~Porod, J.~Romao, and J.~Valle, ``{Neutrino masses and
  mixings from supersymmetry with bilinear R parity violation: A Theory for
  solar and atmospheric neutrino oscillations},'' {\em Phys.Rev.}, vol.~D62,
  p.~113008, 2000, hep-ph/0004115.

\bibitem{Hirsch:2004he}
M.~Hirsch and J.~Valle, ``{Supersymmetric origin of neutrino mass},'' {\em New
  J.Phys.}, vol.~6, p.~76, 2004, hep-ph/0405015.

\bibitem{Bajc:2010qj}
B.~Bajc, T.~Enkhbat, D.~K. Ghosh, G.~Senjanovic, and Y.~Zhang, ``{MSSM in view
  of PAMELA and Fermi-LAT},'' {\em JHEP}, vol.~05, p.~048, 2010, 1002.3631.

\bibitem{Nikolidakis:2007fc}
E.~Nikolidakis and C.~Smith, ``{Minimal Flavor Violation, Seesaw, and
  R-parity},'' {\em Phys.Rev.}, vol.~D77, p.~015021, 2008, 0710.3129.

\bibitem{Csaki:2011ge}
C.~Csaki, Y.~Grossman, and B.~Heidenreich, ``{MFV SUSY: A Natural Theory for
  R-Parity Violation},'' 2011, 1111.1239.
\newblock * Temporary entry *.

\bibitem{Allanach:2003eb}
B.~Allanach, A.~Dedes, and H.~Dreiner, ``{R parity violating minimal
  supergravity model},'' {\em Phys.Rev.}, vol.~D69, p.~115002, 2004,
  hep-ph/0309196.

\bibitem{Ellis:1998rj}
J.~R. Ellis, S.~Lola, and G.~G. Ross, ``{Hierarchies of R violating
  interactions from family symmetries},'' {\em Nucl.Phys.}, vol.~B526,
  pp.~115--135, 1998, hep-ph/9803308.

\bibitem{Giudice:1998bp}
G.~F. Giudice and R.~Rattazzi, ``{Theories with gauge-mediated supersymmetry
  breaking},'' {\em Phys. Rept.}, vol.~322, pp.~419--499, 1999, hep-ph/9801271.

\bibitem{Colangelo:2008qp}
G.~Colangelo, E.~Nikolidakis, and C.~Smith, ``{Supersymmetric models with
  minimal flavour violation and their running},'' {\em Eur.Phys.J.}, vol.~C59,
  pp.~75--98, 2009, 0807.0801.

\bibitem{Minkowski:1977sc}
P.~Minkowski, ``{Mu $\rightarrow$ E Gamma at a Rate of One Out of 1-Billion
  Muon Decays?},'' {\em Phys.Lett.}, vol.~B67, p.~421, 1977.

\bibitem{GellMann:1980vs}
M.~Gell-Mann, P.~Ramond, and R.~Slansky, ``{Complex Spinors and Unified
  Theories},'' pp.~315--321, 1979.
\newblock Published in Supergravity, P. van Nieuwenhuizen $\&$ D.Z. Freedman
  (eds.), North Holland Publ. Co., 1979.

\bibitem{Yanagida:1979as}
T.~Yanagida, ``{Horizontal Symmetry and Masses of Neutrinos},'' 1979.
\newblock Edited by Osamu Sawada and Akio Sugamoto. Tsukuba, Japan, National
  Lab for High Energy Physics, 1979. l09p.

\bibitem{Glashow:1979nm}
S.~Glashow, ``{The Future of Elementary Particle Physics},'' {\em NATO
  Adv.Study Inst.Ser.B Phys.}, vol.~59, p.~687, 1980.

\bibitem{Mohapatra:1979ia}
R.~N. Mohapatra and G.~Senjanovic, ``{Neutrino Mass and Spontaneous Parity
  Violation},'' {\em Phys.Rev.Lett.}, vol.~44, p.~912, 1980.

\bibitem{Schechter:1980gr}
J.~Schechter and J.~Valle, ``{Neutrino Masses in SU(2) x U(1) Theories},'' {\em
  Phys.Rev.}, vol.~D22, p.~2227, 1980.

\bibitem{Schechter:1981cv}
J.~Schechter and J.~Valle, ``{Neutrino Decay and Spontaneous Violation of
  Lepton Number},'' {\em Phys.Rev.}, vol.~D25, p.~774, 1982.

\bibitem{Masina:2002mv}
I.~Masina and C.~A. Savoy, ``{Sleptonarium (constraints on the CP and flavour
  pattern of scalar lepton masses)},'' {\em Nucl. Phys.}, vol.~B661,
  pp.~365--393, 2003, hep-ph/0211283.

\bibitem{Paradisi:2005fk}
P.~Paradisi, ``{Constraints on SUSY lepton flavor violation by rare
  processes},'' {\em JHEP}, vol.~0510, p.~006, 2005, hep-ph/0505046.

\bibitem{Uchiyama:2011zz}
Y.~Uchiyama, ``{Search for lepton flavor violating muon decay: Latest result
  from MEG},'' {\em PoS}, vol.~HQL2010, p.~055, 2011.

\bibitem{Aubert:2009tk}
B.~Aubert {\em et~al.}, ``{Searches for Lepton Flavor Violation in the Decays
  $\tau \rightarrow e \gamma$ and $\tau \rightarrow \mu \gamma$},'' {\em Phys.
  Rev. Lett.}, vol.~104, p.~021802, 2010, 0908.2381.

\bibitem{Hayasaka:2007vc}
K.~Hayasaka {\em et~al.}, ``{New search for $\tau \rightarrow \mu \gamma$ and
  $\tau \rightarrow e \gamma$ decays at Belle},'' {\em Phys. Lett.}, vol.~B666,
  pp.~16--22, 2008, 0705.0650.

\bibitem{Carvalho:2002bq}
D.~Carvalho, M.~Gomez, and J.~Romao, ``{Charged lepton flavor violation in
  supersymmetry with bilinear R-parity violation},'' {\em Phys.Rev.}, vol.~D65,
  p.~093013, 2002, hep-ph/0202054.
\newblock 29 pages, 8 figures. Constraint from solar neutrino data included,
  conclusions changed respect v1.

\bibitem{deCarlos:1996du}
B.~de~Carlos and P.~White, ``{R-parity violation effects through soft
  supersymmetry breaking terms and the renormalization group},'' {\em
  Phys.Rev.}, vol.~D54, pp.~3427--3446, 1996, hep-ph/9602381.

\bibitem{Smirnov:1996bg}
A.~Y. Smirnov and F.~Vissani, ``{Upper bound on all products of R-parity
  violating couplings $\lambda'$ and $\lambda''$ from proton decay},'' {\em
  Phys. Lett.}, vol.~B380, pp.~317--323, 1996, hep-ph/9601387.

\bibitem{Bhattacharyya:1998dt}
G.~Bhattacharyya and P.~B. Pal, ``{New constraints on R-parity violation from
  proton stability},'' {\em Phys. Lett.}, vol.~B439, pp.~81--84, 1998,
  hep-ph/9806214.

\bibitem{Schwetz:2011zk}
T.~Schwetz, M.~Tortola, and J.~W.~F. Valle, ``{Where we are on $\theta_{13}$:
  addendum to 'Global neutrino data and recent reactor fluxes: status of three-
  flavour oscillation parameters'},'' {\em New J. Phys.}, vol.~13, p.~109401,
  2011, 1108.1376.

\bibitem{Nakamura:2010zzi}
K.~Nakamura {\em et~al.}, ``{Review of particle physics},'' {\em J. Phys.},
  vol.~G37, p.~075021, 2010.

\bibitem{Arganda:2005ji}
E.~Arganda and M.~J. Herrero, ``{Testing supersymmetry with lepton flavor
  violating tau and mu decays},'' {\em Phys. Rev.}, vol.~D73, p.~055003, 2006,
  hep-ph/0510405.

\bibitem{Brignole:2004ah}
A.~Brignole and A.~Rossi, ``{Anatomy and phenomenology of mu tau lepton flavour
  violation in the MSSM},'' {\em Nucl. Phys.}, vol.~B701, pp.~3--53, 2004,
  hep-ph/0404211.

\bibitem{Diaz:2003as}
M.~Diaz, M.~Hirsch, W.~Porod, J.~Romao, and J.~Valle, ``{Solar neutrino masses
  and mixing from bilinear R parity broken supersymmetry: Analytical versus
  numerical results},'' {\em Phys.Rev.}, vol.~D68, p.~013009, 2003,
  hep-ph/0302021.

\end{thebibliography}
\bibliographystyle{hieeetr}

\end{document}